\documentclass[12pt]{iopart} 
\usepackage[pdftex]{color,graphicx,hyperref}
\usepackage{bbm}
\usepackage{bm}
\usepackage{verbatim}
\usepackage{wasysym}
\usepackage{ulem}
\usepackage{mathrsfs}
\newcommand{\beq}{\begin{equation}}
\newcommand{\eeq}{\end{equation}}
\newcommand{\ket}[1]{\mbox{$ \mid #1\, \rangle$}}
\newcommand{\bra}[1]{\mbox{$ \langle\, #1\mid$}}

\newcommand{\crea}[2]{\mbox{$\widehat{#1}^{\dagger}_{#2}$}}
\newcommand{\anni}[2]{\mbox{$\widehat{#1}_{#2}$}}

\newcommand{\vecg}[1]{\mbox{${\boldsymbol #1}$}}

\newcommand\unitvec[1]{\widehat{\mathbf{#1}}}
\newcommand{\dert}[2]{\frac{d #1}{d #2}}
\newcommand{\derp}[2]{\frac{\partial #1}{\partial #2}}
\newcommand{\units}[1]{\mbox{#1}}
\newcommand{\text}[1]{\mbox{\footnotesize #1}}
\newcommand{\eqref}[1]{(\ref{#1})}
\newcommand{\email}[1]{\begin{indented}
   						\item[]Email: \texttt{\footnotesize\raggedright#1}
   						\end{indented}}

\begin{document}
\title{The distribution of work performed on a NIS junction}
\author{Jaime E. Santos}
\email{jaie@nanotech.dtu.dk}
\address{Center for Nanostructured Graphene (CNG), Department of Micro and Nanotechnology, 
Technical University of Denmark, DK-2800 Kongens Lyngby, Denmark}
\address{Centro de F\'{i}sica and Departamento de F\'{i}sica, 
Universidade do Minho, P-4710-057 Braga, Portugal}
\author{Pedro Ribeiro}
\email{ribeiro.pedro@gmail.com}
\address{Russian Quantum Center, Novaya Street 100 A, Skolkovo, Moscow Area, 143025 Russia}
\address{CeFEMA, Instituto Superior T\'ecnico, Universidade de Lisboa, Avenida Rovisco Pais, 1049-001 Lisboa, Portugal}
\author{Stefan Kirchner}
\email{stefan.kirchner@correlated-matter.com}
\address{Center for Correlated Matter, Zhejiang University, Hangzhou,  Zhejiang 310058, China}
\address{Max Planck Institute for the Physics of Complex Systems,
N\"othnitzer Str. 38, D-01187 Dresden, Germany}
\address{Max Planck Institute for Chemical Physics of Solids, 
N\"othnitzer Str. 40, D-01187 Dresden, Germany} 

\date{today}
\begin{abstract}
We propose an experimental setup to measure the work performed in a normal-metal/insulator/superconducting (NIS) 
junction, subjected to a voltage change and in contact with a thermal bath. We compute the performed work and argue that the associated heat release can be measured experimentally. Our results are based on an equivalence between the dynamics of the NIS junction and that of an assembly of two-level systems subjected to a circularly polarised field, for which we can determine the work-characteristic function exactly. The  average work dissipated by the NIS junction, as well as its fluctuations, are determined. From the work characteristic function, we also compute the work probability-distribution and show that it does not have a Gaussian character. Our results allow for a direct experimental test of the Crooks-Tasaki fluctuation relation.
\end{abstract}



\maketitle

\section{Introduction}
\label{secInt}

Fluctuation relations are central to our present understanding of statistical mechanics. Their long and distinguished history goes back to at least the studies by Callen and Welton \cite{Callen:1951}, 
Green \cite{Green:1952} and Kubo  \cite{Kubo:1957}, which were inspired by the works of Einstein on the Brownian 
movement \cite{Einstein:1926} and of Johnson \cite{Johnson:1928} and Nyquist \cite{Nyquist:1928} on noise in 
electrical circuits.

The  derivation by Jarzynski of a rather general non-equilibrium work relation  \cite{Jarzynski:1997}, linking the average of the exponential of the work being performed on a system with the equilibrium free energy difference between initial and final equilibrium states of the system, is of particular interest. Subsequently, Crooks \cite{Crooks:1998,Crooks:2000} obtained the Jarzynski equality from a relation between the probability of a given amount of work being performed on a system and the probability that the system performs the same amount of work on its surroundings if the time-reversed protocol is undertaken. Such heightened interest in non-equilibrium work relations has not only been fuelled by novel theoretical advances in out-of-equilibrium dynamics, but also by  experimental progress in preparing and probing non-equilibrium evolution
(see  \cite{Ritort:2003,Ritort:2009,Jarzynski:2010,CampisiRMP:2011,CampisiRMP2:2011} for reviews on the subject).

In the present paper, we propose a relatively simple yet realistic experiment based on a proposal by Crooks \cite{Crooks:2008}, which would allow for a  direct test of the Crooks-Tasaki fluctuation relation. The experimental system we consider consists of a
normal-metal/insulator/superconducting (NIS) junction between a superconductor and a normal metal, which is initially short-circuited and is subjected to a given voltage protocol (see below).
We establish that for the proposed protocol, the full work distribution-function has non-Gaussian character with a non-standard decay exponent. We also compute the first two moments of such distribution, which can be determined experimentally by measuring the average heat released and its variance. 

In order to obtain a closed expression for the characteristic function of the work distribution, we derive an equivalence between the dynamics of an NIS junction and the one of an assembly of two-level systems subjected to a circularly polarized field. Using this mapping, the work distribution is determined. For realistic parameters of experimentally available NIS junctions as {\it e.g.} those of  Ref.~\cite{Lowell:2013}, our findings show that at cryogenic temperatures, even the tiny amounts of heat released will result in a measurable volume change for a probe connected to the junction.

The structure of the paper is as follows: Sect.~\ref{secPer} places the recent developments briefly outlined above into 
their proper historical context and reviews the theoretical tools necessary
to study work-fluctuation relations at the quantum level. In Sect.~\ref{secHm}, we introduce the  Hamiltonian of the NIS junction and describe in detail the work protocol
applied to the system. In Sect.~\ref{secDW}, we use results that are derived in \ref{secMap} and \ref{secgenfun} to compute both the average work dissipated as well as its fluctuations and the full work distribution. \ref{apB} contains an extension of the results presented in \ref{secgenfun}
and \ref{secApTun} contains a discussion of a method of measurement of the tunnelling amplitude that describes the NIS junction.
Section \ref{secIso} contains the main result of our paper, where we compute the numerical value of the average heat released on a NIS junction such as the ones discussed in  Ref.~\cite{Lowell:2013}. We leave to \ref{appIM} the discussion of the experimental techniques that can be used to measure such a quantity. In particular, we show that the heat that is released can be measured by determining the volume change of a probe that absorbs it. We provide estimates of the relative volume change, which should be measurable using capacitive methods~ \cite{White:2002,Pobell:2007}.  Finally, in Sect.~\ref{secConc}, we present our conclusions. For increased readability, most technical details have been relegated to the appendices. 

\section{Perspective and previous works}
\label{secPer}

In this section, we give a brief historical overview of the evolution of
the subject of non-equilibrium fluctuation theorems. Besides the early
contributions already mentioned above, a statement of a general fluctuation theorem involving the free energy of a gas of hard-spheres, interacting at large distances via the Lennard-Jones potential, can be found in the work of Zwanzig \cite{Zwanzig:1954}.
A subsequent important development was the work of Bochkov and  Kuzovlev \cite{Bochkov:1977,Bochkov:1981}, which also concerned the relation of probabilities between a given work protocol and its time reversed version, but which adopted a different perspective to that of Jarzynski and Crooks, namely the former authors considered the generalised force that performs 
work on the system as being external to the dynamical description of the system, 
rather than internal. Another significant contribution  was the work of Evans and Searles (see \cite{Sevick:2008} and 
references therein), which considers the relation between the probability for a dynamical 
trajectory characterised by a certain value of a dissipation function and its time-reversed counterpart. Regarding developments concerning the validity and extension of the Jarzynski-Crooks relations at the classical level, see the experimental works \cite{Hummer:2001, Liphardt:2002,Wang:2002,Collin:2005,Douarche:2005,Schuler:2005,Junier:2009,Mossa:2009,Toyabe:2010,Berut:2012,Saira:2012} and the theoretical ones \cite{Jarzynski:2001,Jarzynski:2004,Cleuren:2006,Palmieri:2007,CampisiPRE:2009,Seifert:2012},
the above list not being exhaustive.

A quantum generalisation  of the Crooks equality was obtained by Tasaki \cite{Tasaki:2000} and also by Kurchan \cite{Kurchan:2000} (for the case of cyclic protocols). A version of the Jarzynski equality valid for quantum systems was also derived by Yukawa \cite{Yukawa:2000}, albeit treating work at a quantum level as an observable, a view that was shown not  to be correct \cite{TalknerPRE:2007}. Such a view lead in the past to a debate concerning the validity of such fluctuation relations at the quantum level \cite{Chernyak:2004,Allahverdyan:2005,Engel:2007}. The extension of the Jarzynski equality to isolated systems, using the definition of the work distribution function of Tasaki and Kurchan, is due to Mukamel \cite{Mukamel:2003}.
A quantum generalisation of the work of Bochkov and Kuzovlev referred above, as well as a clarification of the relation  of their work to that of Jarzynski and Crooks, can be found in a paper by Campisi et al. \cite{Campisi:2011}. 
Extensions of non-equilibrium work relations to isolated systems (micro-canonical fluctuation theorems) are given in \cite{Talkner:2008,Talkner:2013}, whereas work relations valid for arbitrary open quantum systems were obtained in \cite{CampisiPRL:2009}. 
The last reference shows the validity of such theorems for systems that can arbitrarily exchange heat with the bath while a work protocol is being applied to them (see also below). Other theoretical developments concerning quantum systems can be found in the references \cite{Esposito:2006,Talkner:2007,Deffner:2008,Silva:2008,Talkner2:2008,Campisi:2009,Talkner:2009,CampisiPRL:2010,Deffner:2010,Averin:2011,Kawamoto:2011,Yi:2011,
Gambassi:2012,Ngo:2012,Smachia:2012,Smachia:2013,Solinas:2013,Sotiriadis:2013,Marino:2014,Khaymovich:2015,Campisi:2015}, where again the list is not exhaustive. 

At the experimental level, several proposals to demonstrate the validity 
of the Crooks-Tasaki relation in the quantum domain have been presented, including  the use of a series of projective measurements \cite{Huber:2008}, 
the measurement of this relation in the optical spectra of systems subjected to weak quenches \cite{Heyl:2012},
or the use of qubit interferometry \cite{Dorner:2013,Mazzola:2013,Campisi:2013}. The former as well as the latter proposal have been successfully implemented, see \cite{An:2014,Batalhao:2014}. An example of the experimental confirmation of a generalised version of the Jarzynski equality in the quantum domain was discussed in \cite{Khaymovich:2015}. Finally, a discussion of possible solid state experiments performed
on quantum heat engines is given in \cite{Campisi:2015}.

The above mentioned proposal by Crooks \cite{Crooks:2008} relies on an indirect method of observation namely, the measurement of the heat released by a quantum system after a cyclic work protocol has been performed on the system. A specific set-up to measure the heat released by a circuit including a resistor and a Cooper-pair box, based on the temperature increase of a local probe, was already presented in \cite{Pekola:2013}, where the Cooper-pair box acts as a single two-level system. The present proposal has the additional advantage that the heat released scales with the contact area between the normal-metal and the superconducting films.
\begin{figure}
\begin{center} 
\includegraphics[clip,width=12cm]{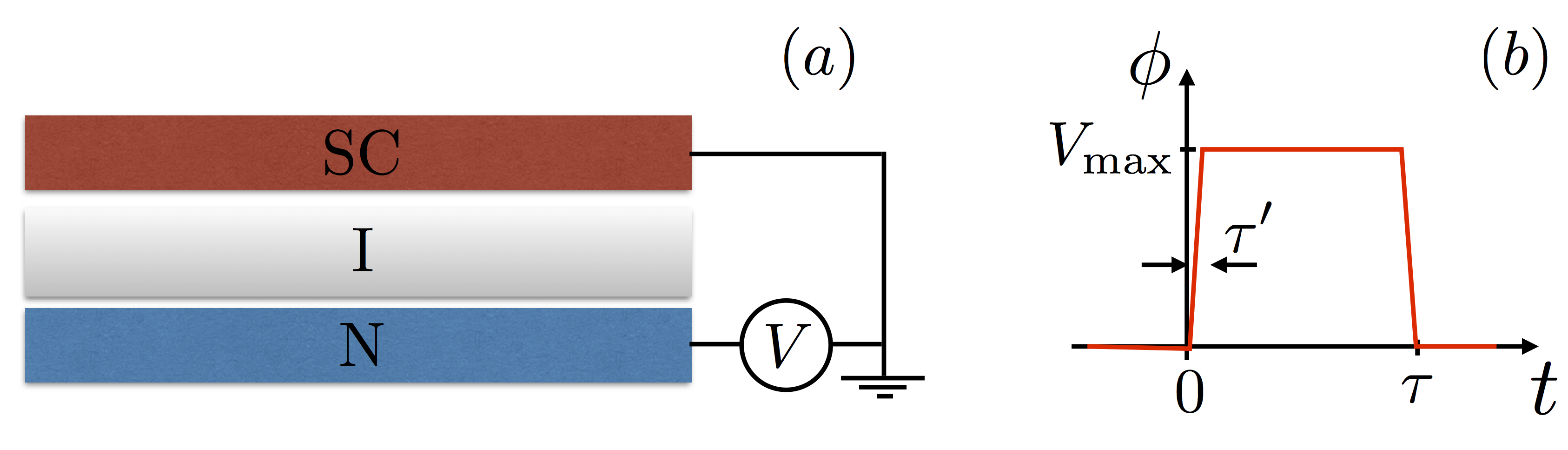}
\par
\end{center}
\caption{(a) NIS junction. (b) Schematic representation of the voltage protocol considered}
\label{fig:1} 
\end{figure}

The work probability distribution, $P_\beta(W,\tau)$, 
is defined, for a class of work protocols where the system is decoupled 
from the thermal bath during the application of the protocol (see below), 
of duration $\tau$, as \cite{Tasaki:2000}
\begin{equation}
P_\beta(W,\tau)=\frac{1}{{\cal Z}(\beta)}\,
\sum_{m,n}\,e^{-\beta E_m^0}\,
\mid\bra{n,\tau}\widehat{U}(\tau,0)\ket{m,0}\mid^2\,\delta(W-E_n^\tau+E_m^0)\,,
\label{PWb}
\end{equation}
where $\ket{m,0}$ are the eigenstates of the system's Hamiltonian at the beginning
of the protocol, 
$\widehat{H}(0)$ (which is given for the NIS junction by \eqref{eqHMc} at $t=0$, see below), and $E_{m}^0$ are the corresponding eigenvalues,
and where $\ket{n,\tau}$ are the eigenstates of the system's Hamiltonian
at the end of the protocol, $\widehat{H}(\tau)$ (which 
is given by the same equation at $t=\tau$), with $E_{n}^\tau$ being the corresponding 
eigenvalues, with $\beta=1/(k_B T)$ and ${\cal Z}(\beta)=Tr(e^{-\beta\widehat{H}(0)})$.
The operator $\widehat{U}(\tau,0)$ is the time-evolution operator of the system
during the protocol, i.e. in the interval $[0,\tau]$, which can be written as a time-ordered exponential of the full Hamiltonian of the system, given for a NIS junction
by eq. \eqref{eqHMc}.

The work characteristic-function of the NIS junction is the Fourier transform of eq.
\eqref{PWb}, and is given by
\begin{equation}
{\cal G}_{\beta}(\upsilon,\tau)=Tr\,\left[\widehat{U}^\dagger(\tau,0)\,e^{i\widehat{H}(\tau)\upsilon}\,
\widehat{U}(\tau,0)\,e^{-i\widehat{H}(0)\upsilon}\,\widehat{\rho}(\beta)\right]\,,
\label{charNIS}
\end{equation}
where $\widehat{\rho}(\beta)=e^{-\beta\widehat{H}(0)}/{\cal Z}(\beta)$
is the density matrix of the junction before the application of the protocol.
 
The unitary evolution of the system in the interval $[0,\tau]$ reflects
the work protocol's restriction mentioned above, i.e. the NIS junction 
is disconnected from the thermal bath at $t=0$ and remains adiabatically insulated 
while the protocol is being applied, being reconnected to the thermal bath at 
$t=\tau$ and undergoing equilibration through the exchange of heat with the bath 
afterwards. It is possible to lift such a restriction on the protocol by explicitly 
considering the interaction of the system with the thermal bath \cite{CampisiPRL:2009}, 
but we will not consider such an extension here, see however the discussion below.

The knowledge of the work  characteristic-function allows one to compute all work 
moments. If one performs the analytic continuation $\upsilon=i\beta$ in the above definition 
and uses the cyclic character of the trace operation, one obtains 
\begin{equation}
{\cal G}_{\beta}(i\beta,\tau)=
{\cal Z}_\tau(\beta)/{\cal Z}(\beta)=e^{-\beta \Delta F}\,,
\label{eqJar}
\end{equation}
where ${\cal Z}_\tau(\beta)=Tr(e^{-\beta\widehat{H}(\tau)})$ is the partition
function for a junction in equilibrium, described by the
Hamiltonian $\widehat{H}(\tau)$, $\Delta F$ being the difference in free-energies
between the final equilibrium state (after thermalisation at a time $t>\tau$)
and the initial equilibrium state at $t=0$. Since in our case the Hamiltonian
goes through a cycle, $\Delta F=0$. Expressing ${\cal G}_{\beta}(i\beta,\tau)$
through its Fourier transform, we obtain 
\begin{equation}
\langle e^{-\beta W}\rangle=
\int_{-\infty}^{+\infty}\,dW\,e^{-\beta W}\,P_\beta(W,\tau)=e^{-\beta \Delta F}=1\,,
\label{eqJarA}
\end{equation}
which is a statement of the Jarzynski equality, which the work probability distribution
satisfies by definition. Considering the relation between ${\cal G}_{\beta}(\upsilon,\tau)$
and ${\cal G}_{\beta}(\upsilon+i\beta,\tau)$, using again the cyclic character of the trace,
it is possible to derive the identity  $P_{\beta}(-W,-\tau)=e^{-\beta W}
P_{\beta}(W,\tau)$ where by $P_{\beta}(W,-\tau)$ we denote the work probability
distribution for the time-reversed protocol. This is a statement of the Crooks-Tasaki 
relation.

Note that in the case of cyclic protocols, one also has $\Delta U=0$, where $U$ denotes the internal
energy of the system. Therefore, from the first law of thermodynamics, $\Delta U=\langle W\rangle_\beta-\Delta Q$,
where $\Delta Q$ denotes the average heat released by the system on which the work protocol is
applied, we obtain $\Delta Q=\langle W\rangle_\beta$, which shows that for cyclic protocols the measurement
of the released heat allows for the determination of $\langle W\rangle_\beta$, as stated.

\section{NIS junction model}
\label{secHm}

The NIS junction is composed of a normal-metal film, a thin insulating layer and a superconducting film. We consider that such a junction can be described by an Hamiltonian consisting of three terms: a Fermi-gas describing the normal-metal film, a BCS superconductor describing the superconducting film and a tunnelling Hamiltonian coupling the two  \cite{Bardeen:1961,Cohen:1962,Ambegaokar:1963,Ambegaokar:1963E,McMillan:1968,Blonder:1982,Cuevas:1996}, which gives rise to induced superconductivity in the normal-metal film through the proximity effect (see \cite{Wolf:2012} and references therein). We use the Hamiltonian
\begin{eqnarray}
\widehat{H}(t)=\widehat{H}_n(t)+\widehat{H}_s+\widehat{X}.
\label{eqHMc}
\end{eqnarray}
The operator
$\widehat{H}_n(t)=\sum_{\mathbf{k},\sigma}\,(\xi_{\mathbf{k}}^n-e\,\phi(t))\,\crea{c}{n\mathbf{k}\sigma}
\anni{c}{n\mathbf{k}\sigma}$ is a free-fermion Hamiltonian 
describing the normal-metal, with $\phi(t)$ being the voltage difference 
across the junction, whose value changes in time. The operator 
\begin{equation}
\widehat{H}_s=\sum_{\mathbf{k},\sigma}\,\xi_{\mathbf{k}}^s\,\crea{c}{s\mathbf{k}\sigma}
\anni{c}{s\mathbf{k}\sigma}+\sum_{\mathbf{k}}\left(\Delta_{\mathbf{k}}\crea{c}{s\mathbf{k}\uparrow}
\crea{c}{s-\mathbf{k}\downarrow}+\overline{\Delta}_{\mathbf{k}}\anni{c}{s-\mathbf{k}\downarrow}
\anni{c}{s\mathbf{k}\uparrow}\right)\,,
\label{eqHSc}
\end{equation}
is the Hamiltonian describing the BCS superconductor, where the pairing function 
$\Delta_{\mathbf{k}}$ is different from zero on a band of width $2\hbar\omega_D$ 
around the Fermi-level, where $\omega_D$ is the superconducting material's Debye frequency. Finally, the third term of Eq.~\eqref{eqHMc} is given by
\begin{equation}
\widehat{X}=\sum_{\mathbf{k},\sigma}\,
t^{\sigma}(\vecg{k})
\left(\crea{c}{s\mathbf{k}\sigma}
\anni{c}{n\mathbf{k}\sigma}+
\crea{c}{n\,-\mathbf{k}\,-\sigma}\anni{c}{s\,-\mathbf{k}\,-\sigma}\right)\,,
\label{eqHX}
\end{equation}
and represents the tunnelling process across the junction. We consider the overall tunnelling matrix element to be small with respect to the gap 
function $\Delta_{\mathbf{k}}$. Such matrix element is taken to be invariant under time-reversal, i.e. 
$\overline{t}^{\sigma}(\vecg{k})=t^{\,-\sigma}(-\vecg{k})$ \cite{McMillan:1968}.
The fermions' momentum is directed along the plane of the films and we assume it to be conserved in a tunnelling process. We will measure the kinetic energy of electrons in the normal-metal film and in the superconducting film with respect to their common Fermi-energy $\mu$.  The super- or subscripts {\it n} and  {\it s} refer to operators or quantities 
pertaining to the normal-metal or to the superconductor, respectively. 
The magnitude of the tunnelling amplitude will depend on the width of
the aluminium oxide layer and its value can be adequately measured for a given device using the method discussed in \ref{secApTun}.
In this method, the power dissipated by the junction is measured at a constant applied voltage equal in value to the energy gap of the superconductor (divided by $e$), which is much
larger than the value used in our protocol.

We consider the NIS to be initially ($t<0$) in equilibrium 
with an heat bath at temperature $T$ and that the voltage $\phi$ across 
the junction is zero. Such voltage is changed from $0$ to a finite value 
$V_{\text{max}}$ and returned to zero within a finite time-interval starting at $t=0$ and
ending at $t=\tau$. During such an interval the junction is decoupled from
the heat bath. For simplicity, we consider that the voltage protocol is symmetric
with respect to the mid-point of the interval $t=\tau/2$. However, the calculation presented
in \ref{secMap} is valid for more general protocols.
Also, $\phi(t)$  varies smoothly within a time-interval $\tau'\ll\tau$ and is returned to zero at the end of the protocol within the same time frame (see figure \ref{fig:1} (b)). 

Physically, for small values of the tunnelling amplitude, the term given by eq. \eqref{eqHX} leads to the opening of
a gap between the energy bands of the normal metal, of magnitude $\Delta^n\ll\Delta$, leaving the energy bands of the superconductor
largely unaffected. This is the proximity effect \cite{Wolf:2012}. In our protocol, we 
place ourselves within a particular adiabatic limit,  i.e. we will consider that the switching-time 
$\tau'\gg \hbar/\Delta$, where $\Delta$ is the value of the gap-function within the superconductor. Thus, the voltage
difference across the junction is held approximately constant within the microscopic 
time-scale of the superconductor and the change of voltage does not induce transitions
which involve the superconducting bands. However, we also 
take $\tau\gg\hbar/\Delta^n\gg\tau'\gg\hbar/\Delta$. This implies that the dynamics is fully diabatic with
respect to the separation between the two energy-bands in the normal metal and thus that the voltage protocol
induces transitions between these bands, leading to work being performed on the system, see \ref{secgenfun}.

Using the Hamiltonian for an NIS junction and the voltage protocol specified, the work probability distribution can be obtained using Eq.~\eqref{charNIS}, 
as will be shown in the next section.
\section{Work dissipated due to the application of the voltage protocol to the junction}
\label{secDW}

In this section, we establish that the dynamics of the NIS junction is equivalent to 
that of an assembly of quantum spins in a time-dependent magnetic field. This entails
an equality between the work probability distributions of both systems, from which it
follows that one can compute the average heat released by the NIS junction, as
well as its higher order fluctuations.

Starting from Eq.~\eqref{eqHMc}, the explicit dependence of the Hamiltonian on the time-dependent bias voltage can be eliminated by a gauge transformation at the expense of acquiring a time-dependent hopping term between the normal and superconducting films. Thus, the Hamiltonian becomes 
\begin{eqnarray}
\widehat{H}_1(t)&=& \widehat{H}_n(0)+ \widehat{H}_s+\widehat{\cal X}(t) 
\label{eqHW_1}
\end{eqnarray}
with 
\begin{eqnarray}
\widehat{\cal X}(t)&=& \sum_{\mathbf{k},\sigma}\,t^{\sigma}(\vecg{k})
\,\left(\,e^{ie\Phi(t)/\hbar}\,\crea{c}{s\mathbf{k}\sigma}
\anni{c}{n\mathbf{k}\sigma}+
e^{-ie\Phi(t)/\hbar}\,\crea{c}{n\,-\mathbf{k}\,-\sigma}\anni{c}{s\,-\mathbf{k}\,-\sigma}\right)\,,
\label{eqHW_2}
\end{eqnarray}
where $\Phi(t)=\int_0^t\,du\,\phi(u)$.
Anticipating that, in general, $t^{\sigma}(\vecg{k}) \ll \Delta$, one can, in the specific adiabatic limit discussed above, 
perform a  generalised Schrieffer-Wolff transformation that takes into account the 
explicit time-dependence of $\widehat{\cal X}(t)$. Furthermore, such a transformation can only be performed
if $eV_{\text{max}}\ll \Delta$, see \ref{secMap}. After such a transformation, 
the metal and superconducting films effectively decouple and only the transformed metallic excitations are subjected to the time-dependent bias. The Hamiltonian, 
$\widehat{\cal H}_n(t)$, which describes the normal-metal film, explicitly reads
\begin{eqnarray}
\widehat{\cal H}_n(t)&=&\sum_{\mathbf{k},\sigma}\,\tilde{\xi}_{\mathbf{k}\sigma}^n\,\crea{c}{n\mathbf{k}\sigma}
\anni{c}{n\mathbf{k}\sigma}\nonumber\\
&&\mbox{}+\sum_{\mathbf{k}}\left(\Delta_{\mathbf{k}}^n\,e^{-2ie\Phi(t)/\hbar}\,\crea{c}{n\mathbf{k}\uparrow}
\crea{c}{n-\mathbf{k}\downarrow}+\overline{\Delta}_{\mathbf{k}}^n\,e^{2ie\Phi(t)/\hbar}\,\anni{c}{n-\mathbf{k}\downarrow}
\anni{c}{n\mathbf{k}\uparrow}\right)\,,
\label{eqHnmod_1}
\end{eqnarray}
where
\begin{eqnarray}
\tilde{\xi}_{\mathbf{k}\sigma}^n&=&\xi_{\mathbf{k}}^n+
\frac{|t^\sigma(\mathbf{k})|^2\,(\xi_{\mathbf{k}}^n+\xi_{\mathbf{k}}^s)}{(\xi_{\mathbf{k}}^n)^2-
(E_{\mathbf{k}}^s)^2}\,,
\label{eqnemod_1}
\\
\Delta_{\mathbf{k}}^n&=&
-\frac{\Delta_{\mathbf{k}}(
\,|t^\uparrow(\mathbf{k})|^2+|t^\downarrow(\mathbf{k})|^2\,)}{2\,[\,(\xi_{\mathbf{k}}^n)^2-(E_{\mathbf{k}}^s)^2\,]}\,,
\label{eqnemod_2}
\end{eqnarray}
are, respectively, the renormalized kinetic energy of electrons
and the induced superconducting parameter, in the normal metal,
due to the proximity effect, $E_{\mathbf{k}}^s=\sqrt{(\xi_{\mathbf{k}}^s)^2+|\Delta_{\mathbf{k}}|^2}$
being the energy of the superconducting excitations. This result is shown in detail in \ref{secMap}.
We thus conclude that for spin-independent tunnelling matrix elements,
the Hamiltonian describing the normal-metal is equivalent, in each 
$\left((\vecg{k},\uparrow),(-\vecg{k},\downarrow)\right)$ subspace, to the
Hamiltonian of a two-level system under the action of a circularly polarised field.
Furthermore, the Hamiltonian that describes the superconductor does not contribute to the work-characteristic function, see \ref{secMap}. 

One can obtain the work characteristic function for a NIS junction simply by considering 
the product of individual characteristic-functions for two-level systems subjected to a 
circularly polarised field, see \ref{secgenfun}.  The logarithm
of the overall work characteristic-function, which is the generating function of the connected 
work-moments, is given by
\begin{eqnarray}
{\cal W}_{\beta}(\upsilon,\tau)&\equiv &
\ln {\cal G}_{\beta}(\upsilon,\tau)
=\sum_{\mathbf{k}}\,\ln\,\Big\{\,
(1-p_\mathbf{k})\nonumber\\
&&\left.\mbox{}+\,p_\mathbf{k}\cdot
\frac{\cosh\left[\,\left(\beta+2i\upsilon\right)\sqrt{|\Delta^n_\mathbf{k}|^2+(\tilde{\xi}_\mathbf{k}^n)^2}\,
\right]}{\cosh\left(\beta\sqrt{|\Delta^n_\mathbf{k}|^2+(\tilde{\xi}_\mathbf{k}^n)^2}\right)}
\right\}\,,
\label{charfAS}
\end{eqnarray}
with $p_\mathbf{k}=\frac{|\Delta^n_\mathbf{k}|^2\,(eV_{\text{max}})^2}{(|\Delta^n_\mathbf{k}|^2+(\tilde{\xi}_\mathbf{k}^n)^2)(|\Delta^n_\mathbf{k}|^2+(\tilde{\xi}_\mathbf{k}^n-eV_{\text{max}})^2)}\,
\sin^2\left(\frac{\tau}{\hbar}\sqrt{|\Delta^n_\mathbf{k}|^2+(\tilde{\xi}_\mathbf{k}^n-eV_{\text{max}})^2}\right)$. The detailed calculation can be found in \ref{secgenfun}.

The average work dissipated per atom in the normal-metal film, $\bar{w}$, 
is obtained from the derivative of Eq. \eqref{charfAS}, evaluated at $\upsilon=0$,
\begin{eqnarray}
\bar{w}&=&\frac{2(eV_{\text{max}})^2}{{\cal N}_{at}}\,\sum_{\mathbf{k}}\,
\tanh\left(\beta\sqrt{|\Delta_{\mathbf{k}}^n|^2+(\tilde{\xi}_{\mathbf{k}}^n)^2}\right)
\nonumber\\
&&\mbox{}\times\,
\frac{|\Delta_{\mathbf{k}}^n|^2\,\sin^2\left(\frac{\tau}{\hbar}\,\sqrt{|\Delta_{\mathbf{k}}^n|^2+(\tilde{\xi}_{\mathbf{k}}^n-eV_{\text{max}})^2}\right)}{\sqrt{|\Delta_{\mathbf{k}}^n|^2+(\tilde{\xi}_{\mathbf{k}}^n)^2}
(|\Delta_{\mathbf{k}}^n|^2+(\tilde{\xi}_{\mathbf{k}}^n-eV_{\text{max}})^2)}
\,,
\label{eqavW1}
\end{eqnarray}
where ${\cal N}_{at}$ is the total number of atoms in the normal-metal film.
From Eq.~\eqref{eqnemod_1}, one has that for weak-coupling between the 
two films, $\tilde{\xi}_{\mathbf{k}}^{n}\approx \xi_{\mathbf{k}}^{n}$. Moreover,
if, as in the system experimentally studied in \cite{Lowell:2013}, the dispersion
relation is the same in both the normal-metal and the superconducting film (since
the normal-metal film is composed of the same material as that 
of the superconducting one, but weakly doped with another metal), we have
from Eq.~\eqref{eqnemod_2} that $\Delta_{\mathbf{k}}^n=\frac{|t(\mathbf{k})|^2}{\Delta_{\mathbf{k}}^*}$.  Assuming that both the order parameter $\Delta_{\mathbf{k}}$ 
and $|t(\mathbf{k})|^2$ only depend on $\mathbf{k}$ through their dependence on
$\xi_{\mathbf{k}}^n$ and are only non-zero in a vicinity of width $\hbar\omega_D$
around the Fermi level of the material composing the films, where $\omega_D$ is the Debye frequency
of the said material, one can transform the above summation into an integral
and write the above equation as
\begin{eqnarray}
\bar{w}&=&2(eV_{\text{max}})^2\,\int_{-\hbar\omega_D}^{\hbar\omega_D}\,d\xi\,\rho(\xi)
\,\tanh\left(\beta\sqrt{|\Delta^n(\xi)|^2+\xi^2}\right)\,
\nonumber\\
&&\mbox{}\times
\frac{|\Delta^n(\xi)|^2\,\sin^2\left(\frac{\tau}{\hbar}\,\sqrt{|\Delta^n(\xi)|^2+(\xi-eV_{\text{max}})^2}\right)}{\sqrt{|\Delta^n(\xi)|^2+\xi^2}\,(|\Delta^n(\xi)|^2+(\xi-eV_{\text{max}})^2)}\,,
\label{eqavW2}
\end{eqnarray}
where we have introduced $\rho(\xi)=\frac{1}{{\cal N}_{at}}\,\sum_{\mathbf{k}}\,
\delta(\xi-\xi_{\mathbf{k}}^n)$, the density of states per atom (and per 
spin-species) in the normal metal film. 

Differentiating Eq.~\eqref{charfAS} twice with respect to $\upsilon$, at
$\upsilon=0$, one obtains the mean-square deviation and thus the relative deviation $\sqrt{\overline{\delta w^2}}/\overline{w}$ of the work performed by the NIS 
junction during the protocol. If one transforms such result into an integral using 
the same assumptions we have used to obtain Eq.~\eqref{eqavW2}, this integral
can also be computed numerically. In a similar fashion, higher moments can be calculated. 

\begin{figure}[ht]
\begin{center} 
\includegraphics[clip,width=12cm]{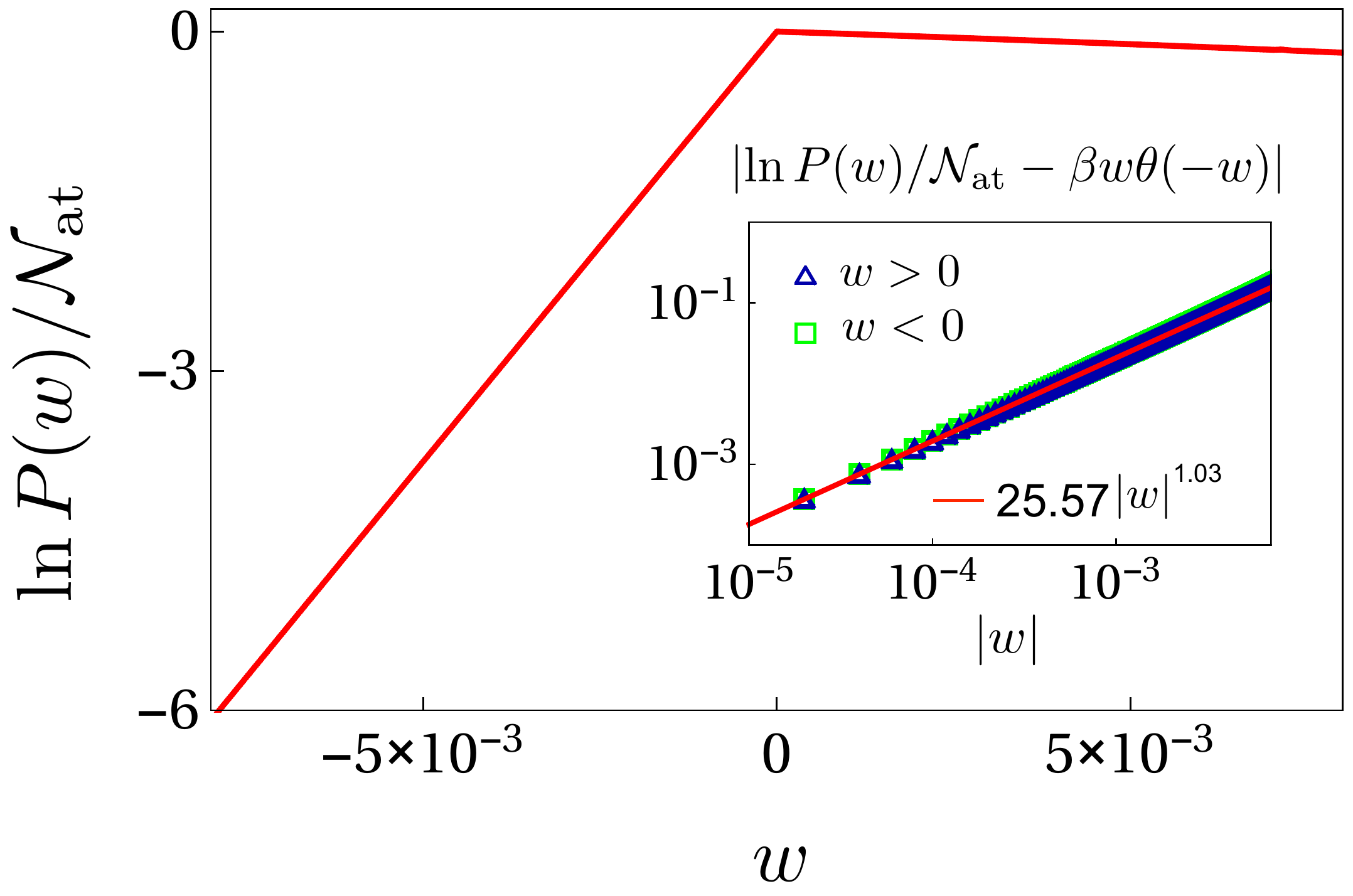} 
\par
\end{center}
\caption{Logarithm of the work per atom distribution for a AlMn/Al$_2$O$_3$/Al junction at $T=1\, K$.}
\label{fig:3} 
\end{figure}

The full characterisation of the work fluctuations requires the determination of the 
work probability distribution, through the computation of the inverse Fourier transform
of ${\cal G}_{\beta}(\upsilon,\tau)$. Writing the probability distribution in terms of 
the logarithm of the characteristic function, as defined in Eq.~\eqref{charfAS}, one has
\begin{equation}
P_\beta(W,\tau)=\int_{-\infty}^{+\infty}\,
\frac{d\upsilon}{2\pi}\,e^{-iW\upsilon\,+{\cal W}_{\beta}(\upsilon,\tau)}\,.
\label{eqPwW}
\end{equation}
Since $W$ is an extensive quantity, {\it i.e.} it is proportional to $\mathcal{N}_{\text{at}}$, the integral can be computed using a saddle-point approximation. Minimising the argument of the exponential in Eq.~\eqref{eqPwW}, one obtains the following saddle-point equation
\begin{equation}
w=\int_{-\hbar\omega_D}^{\hbar\omega_D}\,d\xi\,\rho(\xi)\partial_{x}S(\xi,x)\,,
\label{eqsadd}
\end{equation}
where $w=W/{\cal N}_{at}$ is the work per atom of the sample and $S(\xi,x)$ is
given by
\begin{equation}
S(\xi,x)=\ln\left\{(1-p(\xi))+p(\xi)\cdot
\frac{\cosh\left[\,(\beta+2x)\sqrt{|\Delta^n(\xi)|^2+\xi^2}\,\right]}{\cosh\left[\,\beta\sqrt{|\Delta^n(\xi)|^2+\xi^2}\,\right]}
\right\}\,,
\label{eqS}
\end{equation}
where 
$p(\xi)=\frac{|\Delta^n(\xi)|^2\,(eV_{\text{max}})^2}{(|\Delta^n(\xi)|^2+\xi^2)(|\Delta^n(\xi)|^2+(\xi-eV_{\text{max}})^2)}\,
\sin^2\left(\frac{\tau}{\hbar}\sqrt{|\Delta^n(\xi)|^2+( \xi -eV_{\text{max}})^2}\right)$. Note that the solutions, $x(w)$, of the saddle-point equation \eqref{eqsadd} 
correspond to the analytic continuation of the exponent of the integral given in Eq.~\eqref{eqPwW} to imaginary values of $\upsilon=-ix(w)$.
Numerically solving Eq.~\eqref{eqsadd} for a set of values of $w$ (with the same assumptions as in the two calculations previously performed) and substituting its solution in
Eq.~\eqref{eqPwW}, 
one obtains the result plotted in figure \ref{fig:3} for the logarithm of $P_\beta(W,\tau)$, 
divided by ${\cal N}_{at}$. This calculation indicates that the work probability 
distribution is a stretched exponential for both positive and negative $W$. In the inset, 
the exponent associated with such a stretched exponential is determined for positive $W$
(blue triangles).

We have also subtracted the factor $\beta w$ from the logarithm of the probability distribution at negative $W$
(green squares in the inset) and the function so obtained coincides with the logarithm of the 
probability distribution for positive $W$, thus showing that the distribution obeys the 
Crooks-Tasaki relation at the level of the numerics.

The above figure is a clear indication of the non-Gaussian nature of the work distribution, which may allow the reconstruction of the distribution above from experimental data, using the method that we propose in \ref{appIM}.

\section{Calculation of the dissipated work in the NIS junction}
\label{secIso}

Our primary motivation for the present study was the analysis of the validity of the Crooks-Tasaki relation in an experimentally relevant setting. As discussed above and in \ref{appIM}, this can be accomplished by measuring the heat released in the NIS junction during a full cycle of the protocol. Since such measurements require a high degree of control of the energy flows occurring in an experimental set-up, it is advantageous that the released heat scales with the contact area between the superconducting and normal metal films. 

We now consider the computation of the dissipated work released as heat by the NIS junction.
As a concrete example, we will consider the junction studied in reference  \cite{Lowell:2013}, 
where manganese doped aluminium (the Mn concentration being 0.4\% in mass) acts as the normal-metal electrode in the junction and pure aluminium acts as the superconducting electrode (with aluminium-oxide being the insulating material).

In aluminium, the density of states is approximately constant in the vicinity of the Fermi surface yielding $\rho\approx 0.0117 \omega_D^{-1}$ per atom (and per spin)  \cite{Snow:1967,Smrca:1970}, in units of the Debye frequency, which is approximately $\omega_D = 10^{14} \,\units{Hz}$ in aluminium. This metal is a BCS superconductor with a transition temperature, $T_c=1.20 \,\units{K}$ \cite{Matthias:1963}, and the gap-function
given by $\Delta=1.764\,k_B \,T_c$, where one uses the universal (BCS) relation  \cite{Tinkham:2004}, valid in the weak-coupling limit. For an intermediate coupling superconductor the gap is larger than predicted by the BCS relation. This is of no concern 
here, since we assumed $\tau \gg \tau^{\prime} \gg \hbar/\Delta \simeq 3.6\times 10^{-12} \,\units{s} $. 

For concreteness, we consider the example of a protocol of duration $\tau=10^{-6}\,\units{s}$ with an applied voltage  $V_{\text{max}}=8\times10^{-6}\,\units{V}$ at a temperature of $T=1\,\units{K}$. Performing the integral in \eqref{eqavW2}, assuming a typical tunnelling element $t(\boldsymbol{k}) \simeq 0.05 \Delta$, we obtain an average work 
$\bar{w}\approx 3.2\times 10^{-12}\,\hbar\omega_D$ per atom which amounts to approximately $0.7\,\units{n} \units{J}$ of dissipated work per gram of aluminium. For these parameters, we also obtain a standard deviation of $\overline{\delta w^2}\approx 8.4\times 10^{-14} (\hbar\omega_D)^2$ corresponding to a relative deviation, $\sqrt{\overline{\delta w^2}}/\overline{w}\approx 2.9\times 10^4$, which is a large value for such a quantity. 

The measurement of the released heat can be performed using several techniques, of which the measurement of the volume
change of the film in isothermal conditions seems to be the most promising one, see \ref{appIM} for a detailed discussion.

A word is due with respect to the approximation that the dynamics is unitary during the application of the 
protocol, i.e. that no heat is exchanged with the thermal bath in the time interval $t\in [0,\tau]$, occurring only
at later stages. The exchange of energy between the electronic degrees of freedom and the heat bath occurs mainly through
electron-phonon inelastic collisions. The dependence of resistivity on temperature due to electron-phonon collisions
is given, in metals, by $\rho(T)=\rho_0\,(T/T_0)^5$, where $T_0$ is a temperature of the order of the Debye temperature, 
which is $428\, \units{K}$ in aluminium \cite{Kittel}. Using the value $\rho_0=3.875\times 10^{-8}\,\Omega\,\units{m}$,
with $T_0=400\, \units{K}$ \cite{Desai:1984} for aluminium, and assuming a Drude like dependence of the resistivity on the electron-phonon
relaxation time $\tau_{\text{\tiny pe}}$, $\rho(T)=\frac{m^*}{ne^2\tau_{\text{\tiny pe}}(T)}$, where $m^*=1.10\,m_e$ is the electron effective
mass in aluminium \cite{Levinson:1983} and $n=2.1\times 10^{29}\,\units{m}^{-3}$ is the electronic density in aluminium, we
obtain $\tau_{\text{\tiny pe}}(T=1\,\units{K})\approx 5\times 10^{-2} \units{s}$, which is much larger than the protocol duration $\tau$. Hence,
the said approximation is fully justified. Moreover, such relaxation time is of the order of the equilibration time that is necessary to wait after the 
end of the protocol before the volume deviation discussed in \ref{appIM} can be measured.

\section{Conclusions}

Fluctuation-dissipation relations play a central role in thermodynamics and the approach to equilibrium, but the direct experimental verification of the various relations can often be difficult. In this work, we discussed a simple but realistic experimental set-up that will allow to test the Crooks-Tasaki fluctuation relation.  Our proposal consists of an NIS junction of a superconductor and a normal metal where the bias voltage is altered according to a specified protocol, see Fig.~\ref{fig:1}-(b).

As shown, the released heat scales with the contact area between the superconducting and normal metal films. 
Our proposal is thus amenable to experimental verification, e.g. by coupling the junction to a probe that isothermally absorbs the released heat and changes its volume as a result. Order of magnitude estimates of this effect for the aluminium-based junctions of Ref. \cite{Lowell:2013} were presented in\ref{appIM}. 
 
The calculation of the full work distribution function and its first moments clearly establishes an exponential behavior of such quantity that displays a non-Gaussian character with a non-standard decay exponent. 

Our results are based on an equivalence between the dynamics of a NIS junction 
and the dynamics of an assembly of two-level systems subjected to a circularly polarized field. 
This equivalence holds in a particular adiabatic limit in which the
switching-on/off of the potential difference across the junction takes place in time scales much larger than the microscopic time scale of the system, determined by the value of the superconducting gap. 

\label{secConc}

\textbf{Acknowledgements:}  We acknowledge helpful discussions with O. Stockert,
J. Pekola, E. Lutz, M. Kiselev, A. Muramatsu, K. Kroy, E. Lage, J. L. Santos, M. I. Vasilevskiy, 
N. M. R. Peres and A.-P. Jauho. J.\,E.\,Santos work contract at the University of Minho, where most of
the present research was conducted,  was financed in the framework of the Program of 
Recruitment of Post Doctoral Researchers for the Portuguese Scientific 
and Technological System, within the Operational Program Human Potential (POPH) of the QREN, 
participated by the European Social Fund (ESF) and national funds of the 
Portuguese Ministry of Education and Science (MEC). He also acknowledges 
support provided to the current research project by FEDER through the COMPETE 
Program and by FCT in the framework of the Strategic Project PEST-C/FIS/UI607/2011.
He would also like to thank the support of the MPIPkS in Dresden within the framework
of their  Visitors' Program at several stages of this work, as well as the support of the Center for Nanostructured Graphene (CNG) 
at the Technical University of Denmark at the final stage of the work. The CNG is sponsored by the Danish National 
Research Foundation, Project No. DNRF58. P.\,Ribeiro acknowledges financial support from FCT through the contract Ref. IF/00347/2014/CP1214/CT0002 under the IF2014 program. 
S.\,Kirchner acknowledges partial support by the National Natural Science Foundation of China, grant No.11474250.


\appendix
\section{Mapping of the work characteristic-function to that of an assembly of
two-level systems}
\label{secMap}

Following \cite{Rogovin:1974} and starting from the Hamiltonian \eqref{eqHMc}, we
change representation, so as to write the original time-evolution operator $\widehat{U}(t,0)$,
determined by such Hamiltonian, in terms of the time-evolution operator in a 
new representation, $\widehat{V}(t,0)$, 
with $0\leq t\leq\tau$, 
\begin{equation}
\widehat{U}(t,0)=
e^{ie\Phi(t)\widehat{N}_n/\hbar}\,\widehat{V}(t,0)\,,
\label{eqUV}
\end{equation}
where $\Phi(t)=\int_0^t du\,\phi(u)$ and $\widehat{N}_n=\sum_{\mathbf{k},\sigma}\,
\crea{c}{n\mathbf{k}\sigma}\anni{c}{n\mathbf{k}\sigma}$ is the number of particles
on the normal side of the junction. This change of representation corresponds 
to a gauge transformation that eliminates the dependence on $\phi(t)$ of $\widehat{H}_n(t)$ and transfers
such dependence to the tunnelling part of the Hamiltonian. The transformed
operator, $\widehat{H}_1(t)=e^{-ie\Phi(t)\widehat{N}_n/\hbar}\widehat{H}(t)
e^{ie\Phi(t)\widehat{N}_n/\hbar}$, is given by $\widehat{H}_1(t)=\widehat{H}_0+\lambda\,\widehat{\cal X}(t)$, where $\lambda$ is a small dimensionless parameter 
that sets the scale of the tunnelling matrix element and which is explicitly written here
for convenience of calculation. The operator $\widehat{H}_0=\widehat{H}_n(0)+\widehat{H}_s$ describes the  normal-metal film and the superconducting film at $t=0$, 
in the absence of the tunnelling operator
\eqref{eqHX}. On the other hand, the operator $\widehat{\cal X}(t)$ is given by eq. \eqref{eqHX}.

The operator $\widehat{V}(t,0)$ can be expressed in terms of a 
time-ordered  exponential involving the transformed Hamiltonian 
$\widehat{H}_1(t)$. This exponential can be written in terms of 
an ordered product of exponentials, evaluated at increasing times, as
\begin{equation}
\widehat{V}(t,0)=\prod_{l=1}^{N}\,\exp\left[-\frac{it}{\hbar N}\,
\widehat{H}_1(u_l)\,\right]\,
\label{eqTrott}
\end{equation}
where $N\rightarrow\infty$ and $u_l=l\cdot t/N$. We now introduce a second-unitary
transformation, dependent on both $\lambda$ and on $t$, by defining a transformed
Hamiltonian operator as 
\begin{equation}
\widehat{H}_2(t)=e^{i\lambda\widehat{S}(t)}\,\widehat{H}_1(t)\,e^{-i\lambda\widehat{S}(t)}\,,
\label{eqUnit}
\end{equation}
where the Hermitian operator $\widehat{S}(t)$ will be adequately chosen below. Substituting
this result in \eqref{eqTrott}, we obtain
\begin{equation}
\widehat{V}(t,0)=e^{-i\lambda\widehat{S}(t)}\,\prod_{l=1}^{N}\,\left[\exp\left[-\frac{it}{\hbar N}\,
\widehat{H}_2(u_l)\right]\,e^{i\lambda\widehat{S}(u_l)}\,e^{-i\lambda\widehat{S}(u_{l-1})}
\right]\,e^{i\lambda\widehat{S}(0)}\,.
\label{eqTrott1}
\end{equation}
Using the Baker-Campbell-Hausdorf formula, one can write, to order $\lambda^2$,
and to order $1/N$
\begin{eqnarray}
e^{i\lambda\widehat{S}(u_l)}\,e^{-i\lambda\widehat{S}(u_{l-1})}&\approx&
e^{i\lambda(\widehat{S}(u_l)-\widehat{S}(u_{l-1}))+\frac{\,\lambda^2}{2}
\,[\widehat{S}(u_l),\widehat{S}(u_{l-1})]_{-}}\nonumber\\
&\approx&e^{\frac{i\lambda}{N}\,\left(\dert{\widehat{S}}{u}+\frac{i\lambda}{2}\left[\widehat{S}(u_l),\dert{\widehat{S}}{u}\right]_{-}
\right)}\,,
\label{eqTrott2}
\end{eqnarray}
where $[\;,\;]_{-}$ is the commutator of two operators and where the derivative of $\widehat{S}(u)$
is computed at $u_l$. Substituting \eqref{eqTrott2} in \eqref{eqTrott1}, we obtain to order 
$\lambda^2$ and order $1/N$, the result
\begin{equation}
\widehat{V}(t,0)=e^{-i\lambda\widehat{S}(t)}\,\prod_{l=1}^{N}\,\exp\left[-\frac{it}{\hbar N}\,
\widehat{\cal H}(u_l)\right]\,e^{i\lambda\widehat{S}(0)}\,,
\label{eqTrott3}
\end{equation}
where, to order $\lambda^2$, the operator $\widehat{\cal H}(t)$ is given by
\begin{eqnarray}
\widehat{\cal H}(t)&\approx&e^{i\lambda\widehat{S}(t)}\,
(\,\widehat{H}_0+\lambda\,\widehat{\cal X}(t)\,)\,e^{-i\lambda\widehat{S}(t)}-\lambda\hbar\dert{\widehat{S}}{t}
-\frac{i\hbar\lambda^2}{2}\left[\,\widehat{S}(t),\dert{\widehat{S}}{t}\,\right]_{-}\nonumber\\
&\approx&
\widehat{H}_0+\lambda\left(-\hbar\dert{\widehat{S}}{t}+\widehat{\cal X}(t)+i\left[\,\widehat{S}(t),
\widehat{H}_0\,\right]_{-}\right)\nonumber\\
&&\mbox{}+\frac{i\lambda^2}{2}
\left[\,\widehat{S}(t)\,,\,-\hbar\dert{\widehat{S}}{t}+\widehat{\cal X}(t)+i\left[\widehat{S}(t),
\widehat{H}_0\,\right]_{-}\,\right]_{-}\nonumber\\
&&\mbox{}+\frac{i\lambda^2}{2}
\left[\,\widehat{S}(t)\,,\,\widehat{\cal X}(t)\,\right]_{-}\,.
\label{eqtransf}
\end{eqnarray}

We choose the hermitian operator $\widehat{S}(t)$ such that the linear term in $\lambda$ vanishes
in \eqref{eqtransf}, i.e. such operator obeys the first order ordinary differential equation
\begin{equation}
\hbar\dert{\widehat{S}}{t}=\widehat{\cal X}(t)+i\left[\,\widehat{S}(t),
\widehat{H}_0\,\right]_{-}\,,
\label{eqdiffS}
\end{equation}
whereas the transformed Hamiltonian reduces to
\begin{equation}
\widehat{\cal H}(t)=\widehat{H}_0+\frac{i\lambda^2}{2}
\left[\,\widehat{S}(t)\,,\,\widehat{\cal X}(t)\,\right]_{-}\,.
\label{eqSWt}
\end{equation}
One still needs to specify an initial condition for \eqref{eqdiffS} to be 
properly defined, which we will do below. The above choice of $\widehat{S}(t)$
corresponds to a time-dependent Schrieffer-Wolff transformation \cite{Foldy:1950,Schrieffer:1966},
which is reminiscent of the work of \cite{Wegner:1994,Kehrein:1996}. One
can thus write the operator $\widehat{U}(t,0)$ as $\widehat{U}(t,0)=e^{ie\Phi(t)\widehat{N}_n/\hbar}\,e^{-i\lambda\widehat{S}(t)}\,
\widehat{\mathscr{V}}(t,0)\,e^{i\lambda\widehat{S}(0)}$, where 
\begin{equation}
\widehat{\mathscr{V}}(t,0)=T\,\exp\left[-\frac{i}{\hbar} \int_{0}^{t}\,du\,
\widehat{\cal H}(u)\,\right]\,,
\label{eqVtransf}
\end{equation}
which is valid to order $\lambda^2$ and where $\widehat{\cal H}(u)$ is given by \eqref{eqSWt}.
Substituting this expression for $\widehat{U}(t,0)$ in \eqref{charNIS} and using the cyclic
character of the trace operation, one obtains
\begin{eqnarray}
{\cal G}_{\beta}(\upsilon,\tau)&=&Tr\,\left[\widehat{\mathscr{V}}^\dagger(\tau,0)\,e^{i\lambda\widehat{S}(\tau)}\,
e^{i\widehat{H}_1(\tau)\upsilon}\,e^{-i\lambda\widehat{S}(\tau)}\right.\nonumber\\
&&\mbox{}\left.\cdot\,\widehat{\mathscr{V}}(\tau,0)\,e^{i\lambda\widehat{S}(0)}\,\,e^{-i\widehat{H}(0)\upsilon}\,\widehat{\rho}(\beta)
\,e^{-i\lambda\widehat{S}(0)}\,\right]\,.
\label{charNIS2}
\end{eqnarray}
Under the action of, respectively, $\widehat{S}(0)$ and $\widehat{S}(\tau)$, $\widehat{H}(0)$ 
and $\widehat{H}_1(\tau)$ do not transform into $\widehat{\cal H}(0)$ and $\widehat{\cal H}(\tau)$,
as the factors that involve the derivative of $\hat{S}(t)$ at these points, and which appear
in \eqref{eqtransf}, are not present in the transformation. 
However, we may still fix the derivative of $\hat{S}(t)$
at one of these points, by the choice of an appropriate initial condition. 
We thus choose $\left.\frac{d\widehat{S}}{dt}\right|_{t=0}=0$.
The explicit computation of ${\cal G}_{\beta}(\upsilon,\tau)$ now 
requires the solution of \eqref{eqdiffS}, subjected
to the condition that we have imposed at $t=0$. In order to solve
this equation, we write $\widehat{S}(t)$ in the form
\begin{eqnarray}
\widehat{S}(t)&=&\sum_{k\sigma}[\,r_{\mathbf{k}}^\sigma(t)\,\crea{c}{s\mathbf{k}\sigma}
\anni{c}{n\mathbf{k}\sigma}+\overline{r}_{\mathbf{k}}^\sigma(t)\,\crea{c}{n\mathbf{k}\sigma}
\anni{c}{s\mathbf{k}\sigma}\,]\nonumber\\
&&\mbox{}+\sum_{k\sigma}[\,\overline{s}_{\mathbf{k}}^\sigma(t)\,\crea{c}{s-\mathbf{k}-\sigma}
\crea{c}{n-\mathbf{k}-\sigma}+s_{\mathbf{k}}^\sigma(t)\,\anni{c}{n-\mathbf{k}-\sigma}
\anni{c}{s-\mathbf{k}-\sigma}\,]\,,
\label{eqansatz}
\end{eqnarray}
where $r_{\mathbf{k}}^\sigma(t)$ and $s_{\mathbf{k}}^\sigma(t)$ are 
complex-valued functions, to be determined. 
Substituting \eqref{eqansatz} in \eqref{eqdiffS} and 
equating the terms pertaining to the same pairs of operators, we obtain the coupled 
system of first-order ordinary differential equations
\begin{equation}
\left\{
\mbox{
\begin{tabular}{l}   
$\dert{r_{\mathbf{k}}^\sigma}{t}=\frac{t^\sigma(\mathbf{k})}{\hbar}\, e^{ie\Phi(t)/\hbar}
+\frac{i}{\hbar}\,(\xi_{\mathbf{k}}^n-\xi_{\mathbf{k}}^s)\,r_{\mathbf{k}}^\sigma+\frac{i}{\hbar}\,\sigma\,\Delta_{\mathbf{k}}\,s_{\mathbf{k}}^\sigma$\\
\\
$\dert{s_{\mathbf{k}}^\sigma}{t}=\frac{i}{\hbar}\,(\xi_{\mathbf{k}}^n+\xi_{\mathbf{k}}^s)\,s_{\mathbf{k}}^\sigma+\frac{i}{\hbar}\,\sigma\,\overline{\Delta}_{\mathbf{k}}\,r_{\mathbf{k}}^\sigma$
\end{tabular}
}\right.\,,
\label{eqSdiff}
\end{equation}
with the initial conditions $\left.\dert{r_{\mathbf{k}}^{\sigma}}{t}\right|_{t=0}
=\left.\dert{s_{\mathbf{k}}^{\sigma}}{t}\right|_{t=0}=0$. These conditions, once
substituted in \eqref{eqSdiff} at $t=0$, yield for $r_{\mathbf{k}}^\sigma(0)$ and
$s_{\mathbf{k}}^\sigma(0)$, the result
\begin{equation}
\left\{
\mbox{
\begin{tabular}{l}   
$r_{\mathbf{k}}^\sigma(0)=\frac{it^\sigma(\mathbf{k})\,(\xi_{\mathbf{k}}^n+\xi_{\mathbf{k}}^s)}{(\xi_{\mathbf{k}}^n)^2-(E_{\mathbf{k}}^s)^2}$\\
\\
$s_{\mathbf{k}}^\sigma(0)=-\frac{it^\sigma(\mathbf{k})\,\sigma\,\overline{\Delta}_{\mathbf{k}}}{(\xi_{\mathbf{k}}^n)^2-
(E_{\mathbf{k}}^s)^2}$
\end{tabular}
}\right.\,.
\label{eqSdiff2}
\end{equation}
This choice of parameters corresponds to the time-independent Schrieffer-Wolff transformation introduced 
in  \cite{Donabidowicz:2008} and such transformation can be applied to the Hamiltonian of the system at 
$t\leq 0$, since $\phi(t)=0$ in this interval.

Writing the pairing function in terms of an amplitude and a phase,
i.e. $\Delta_{\mathbf{k}}=|\Delta_{\mathbf{k}}|\,e^{i\varphi}$, where $\varphi$ is
independent of $\mathbf{k}$, we define the new variables $x_{\mathbf{k}}^{\sigma}(t)=
u_{\mathbf{k}}^{\sigma}(t)\,e^{-\frac{i}{\hbar}\xi_{\mathbf{k}}^nt-\frac{i}{2}\varphi}$,
$y_{\mathbf{k}}^{\sigma}(t)=
v_{\mathbf{k}}^{\sigma}(t)\,e^{-\frac{i}{\hbar}\xi_{\mathbf{k}}^nt+\frac{i}{2}\varphi}$.
Substituting these expressions in eq. \eqref{eqSdiff}
and differentiating the result, we obtain two decoupled second-order equations
for $x_{\mathbf{k}}^{\sigma}(t)$ and $y_{\mathbf{k}}^{\sigma}(t)$, with appropriate
boundary conditions. These equations can be easily solved.

The solutions of such equations, expressed in terms of the original functions $r_{\mathbf{k}}^\sigma(t)$,
$s_{\mathbf{k}}^\sigma(t)$, are given by
\begin{eqnarray}
r_{\mathbf{k}}^\sigma(t)&=&it^\sigma(\mathbf{k})\,e^{\frac{i}{\hbar}\xi_{\mathbf{k}}^nt}\,
\left[\,\frac{\xi_{\mathbf{k}}^n+\xi_{\mathbf{k}}^s}{(\xi_{\mathbf{k}}^n)^2-
(E_{\mathbf{k}}^s)^2}\,\left(\cos(\omega_{\mathbf{k}}^st)-\frac{i\xi_{\mathbf{k}}^n}{E_{\mathbf{k}}^s}
\cdot\sin(\omega_{\mathbf{k}}^st)\right)\right.\nonumber\\
&&\mbox{}+\frac{1}{\hbar E_{\mathbf{k}}^s}\,
\int_0^t du\,(\,e\phi(u)-\xi_{\mathbf{k}}^n-\xi_{\mathbf{k}}^s\,)\nonumber\\
&&\mbox{}\left.\cdot\,e^{\frac{i}{\hbar}(e\Phi(u)-\xi_{\mathbf{k}}^nu)}\,\sin(\omega_{\mathbf{k}}^s(t-u))\,\right]\,,
\label{eqSdiff7}
\end{eqnarray}
and
\begin{eqnarray}
s_{\mathbf{k}}^\sigma(t)&=&-i\sigma\,\overline{\Delta}_{\mathbf{k}}\,
t^\sigma(\mathbf{k})\,e^{\frac{i}{\hbar}\xi_{\mathbf{k}}^nt}\,
\left[\,\frac{1}{(\xi_{\mathbf{k}}^n)^2-
(E_{\mathbf{k}}^s)^2}\,\left(\cos(\omega_{\mathbf{k}}^st)-\frac{i\xi_{\mathbf{k}}^n}{E_{\mathbf{k}}^s}
\cdot\sin(\omega_{\mathbf{k}}^st)\right)\right.\nonumber\\
&&\mbox{}-\left.\frac{1}{\hbar E_{\mathbf{k}}^s}\,
\int_0^t du
\,e^{\frac{i}{\hbar}(e\Phi(u)-\xi_{\mathbf{k}}^nu)}\,\sin(\omega_{\mathbf{k}}^s(t-u))\,\right]\,.
\label{eqSdiff8}
\end{eqnarray}
Now, one can show, using the know properties of the exponential function and integration by parts,
the following identity
\begin{eqnarray}
\frac{1}{\hbar E_{\mathbf{k}}^s}\,
\int_0^t du\,
\,e^{\frac{i}{\hbar}(e\Phi(u)-\xi_{\mathbf{k}}^nu)}\,\sin(\omega_{\mathbf{k}}^s(t-u))
&=&-\frac{1}{(\xi_{\mathbf{k}}^n)^2-
(E_{\mathbf{k}}^s)^2}\,\cdot\,\Big[\,e^{\frac{i}{\hbar}(e\Phi(t)
-\xi_{\mathbf{k}}^nt)}\nonumber\\
&&\mbox{}-\left(\cos(\omega_{\mathbf{k}}^st)-\frac{i\xi_{\mathbf{k}}^n}{E_{\mathbf{k}}^s}
\cdot\sin(\omega_{\mathbf{k}}^st)\right)
\nonumber\\
&&\mbox{}-\frac{i}{\hbar}\,\int_0^t du\,e\phi(u)
\,e^{\frac{i}{\hbar}(e\Phi(u)-\xi_{\mathbf{k}}^nu)}\nonumber\\
&&\mbox{}\cdot\,\Big(\,\cos(\omega_{\mathbf{k}}^s(t-u))
\label{eqidint}
\\
&&\mbox{}-\left.\left.\,\frac{i\xi_{\mathbf{k}}^n}{E_{\mathbf{k}}^s}\cdot
\sin(\omega_{\mathbf{k}}^s(t-u))\right)\right]\,.\nonumber
\end{eqnarray}
Using \eqref{eqidint} in \eqref{eqSdiff7} and \eqref{eqSdiff8}, one can write 
the solution of the system given in \eqref{eqSdiff} as
\begin{eqnarray}
r_{\mathbf{k}}^\sigma(t)&=&\frac{i\,t^\sigma(\mathbf{k})\,(\xi_{\mathbf{k}}^n+\xi_{\mathbf{k}}^s)}{(\xi_{\mathbf{k}}^n)^2-
(E_{\mathbf{k}}^s)^2}\cdot e^{\frac{ie}{\hbar}\Phi(t)}
+\frac{t^\sigma(\mathbf{k})}{\hbar[(\xi_{\mathbf{k}}^n)^2-
(E_{\mathbf{k}}^s)^2]}
\,\int_0^t du\,e\phi(u)
\nonumber\\
&&\mbox{}\times\,e^{\frac{i}{\hbar}(e\Phi(u)+\xi_{\mathbf{k}}^n(t-u))}\cdot\,
\left((\xi_{\mathbf{k}}^n+\xi_{\mathbf{k}}^s)\cdot\cos(\omega_{\mathbf{k}}^s(t-u))\right.
\label{eqSdiff9}
\nonumber\\
&&\mbox{}-\left.i\,\frac{\xi_{\mathbf{k}}^s(\xi_{\mathbf{k}}^n+\xi_{\mathbf{k}}^s)+
|\Delta_{\mathbf{k}}|^2}{E_{\mathbf{k}}^s}\cdot
\sin(\omega_{\mathbf{k}}^s(t-u))\right)\,,
\end{eqnarray}
and
\begin{eqnarray}
\label{eqSdiff10}
s_{\mathbf{k}}^\sigma(t)&=&-\frac{i\,t^\sigma(\mathbf{k})\,\sigma\,\overline{\Delta}_{\mathbf{k}}}{(\xi_{\mathbf{k}}^n)^2-(E_{\mathbf{k}}^s)^2}\cdot e^{\frac{ie}{\hbar}\Phi(t)}
-\frac{t^\sigma(\mathbf{k})\,\sigma\,\overline{\Delta}_{\mathbf{k}}}{\hbar
[(\xi_{\mathbf{k}}^n)^2-(E_{\mathbf{k}}^s)^2]}
\,\int_0^t du\,e\phi(u)\\
&&\mbox{}\times e^{\frac{i}{\hbar}(e\Phi(u)+\xi_{\mathbf{k}}^n(t-u))}\cdot
\left(\cos(\omega_{\mathbf{k}}^s(t-u))-\frac{i\xi_{\mathbf{k}}^n}{E_{\mathbf{k}}^s}\cdot
\sin(\omega_{\mathbf{k}}^s(t-u))\right)\,.\nonumber
\end{eqnarray}
We stated above that we are considering protocols in which the applied voltage difference 
$\phi(t)$ changes from $0$ to $V_{\text{max}}$ in a time interval $\tau'$ i.e. $\phi(t)=V_{\text{max}}\psi(t/\tau')$,
where $\psi(x)$ changes from $0$ to $1$ in an interval of length $1$. The integrals
which enter the second term of both \eqref{eqSdiff9} and \eqref{eqSdiff10} are all, after 
the change of variables $u\rightarrow u/\tau'$, of the form 
\begin{equation}
\frac{eV_{\text{max}}\tau'}{\hbar}\,\int_0^{t/\tau'} du\,\psi(u)\,e^{\frac{ieV_{\text{max}}\tau'}{\hbar}\Psi(u)}
e^{-\frac{i\tau'}{\hbar}(\xi_{\mathbf{k}}^n\pm
E_{\mathbf{k}}^s)u}\,,
\label{intRL1}
\end{equation}
where $\Psi(u)=\int_{0}^u\,dv\,\psi(v)$. The term $\psi(u)\,e^{\frac{ieV_{\text{max}}\tau'}{\hbar}\Psi(u)}$
is a bounded function, whereas the exponential $e^{-\frac{i\tau'}{\hbar}(\xi_{\mathbf{k}}^n\pm
E_{\mathbf{k}}^s)u}$ is at least equal to $e^{-\frac{i\tau'}{\hbar}|\Delta_{\mathbf{k}}|u}$, which in the
adiabatic regime $\tau'\gg\hbar/\Delta$ oscillates rapidly. Thus, provided that $t\gg \tau'$ and 
$eV_{\text{max}}\tau'/\hbar\leq 1$, then \eqref{intRL1} is zero by the Riemann-Lebesgue lemma. From these
conditions, we also obtain $eV_{\text{max}}\ll \Delta$, i.e. the tunnelling voltage applied should be well 
within the gap. Therefore, we conclude that in the adiabatic limit and for $t\gg \tau'$,
\eqref{eqSdiff9} and \eqref{eqSdiff10} reduce to
\begin{equation}
\left\{
\mbox{
\begin{tabular}{l}   
$r_{\mathbf{k}}^\sigma(t)\approx\frac{i\,t^\sigma(\mathbf{k})\,(\xi_{\mathbf{k}}^n+\xi_{\mathbf{k}}^s)}{(\xi_{\mathbf{k}}^n)^2-
(E_{\mathbf{k}}^s)^2}\cdot e^{\frac{ie}{\hbar}\Phi(t)}$\\
\\
$s_{\mathbf{k}}^\sigma(t)\approx-\frac{i\,t^\sigma(\mathbf{k})\,\sigma\,\overline{\Delta}_{\mathbf{k}}}{(\xi_{\mathbf{k}}^n)^2-(E_{\mathbf{k}}^s)^2}\cdot e^{\frac{ie}{\hbar}\Phi(t)}$
\end{tabular}
}\right.\,.
\label{eqSdiff11}
\end{equation}
From such a discussion and since $\phi(\tau)=0$, it also follows that  $\left.\dert{r_{\mathbf{k}}^{\sigma}}{t}\right|_{t=\tau}
\approx\left.\dert{s_{\mathbf{k}}^{\sigma}}{t}\right|_{t=\tau}\approx 0$ and hence we obtain
$\left.\dert{\widehat{S}}{t}\right|_{t=\tau}=0$. In such a case, we can write eq. \eqref{charNIS2}
as
\begin{equation}
{\cal G}_{\beta}(\upsilon,\tau)=Tr\,\left[\widehat{\mathscr{V}}^\dagger(\tau,0)\,
e^{i\widehat{\cal H}(\tau)\upsilon}\,
\widehat{\mathscr{V}}(\tau,0)\,e^{-i\widehat{\cal H}(0)\upsilon}\,\widehat{\rho}(\beta)\,\right]\,,
\label{charNIS4}
\end{equation}
where $\widehat{\cal H}(\tau)$ is given by \eqref{eqSWt} (with $t$ substituted by $\tau$ in that equation). 
Substituting the operator $\widehat{S}(t)$ by its expression as given in \eqref{eqansatz}, with $r_{\mathbf{k}}^\sigma(t)$ and $s_{\mathbf{k}}^\sigma(t)$ given by \eqref{eqSdiff11}, in \eqref{eqSWt}, we
obtain that $\widehat{\cal H}(t)=\widehat{\cal H}_s+\widehat{\cal H}_n(t)$, where the operators
$\widehat{\cal H}_s$ and $\widehat{\cal H}_n(t)$, pertaining, respectively, to the
superconductor and the normal-metal, are given by
\begin{eqnarray}
\widehat{\cal H}_s&=&\sum_{\mathbf{k},\sigma}\,\tilde{\xi}_{\mathbf{k}\sigma}^s\,\crea{c}{s\mathbf{k}\sigma}
\anni{c}{s\mathbf{k}\sigma}
+\sum_{\mathbf{k}}\left(\Delta_{\mathbf{k}}^s\crea{c}{s\mathbf{k}\uparrow}
\crea{c}{s-\mathbf{k}\downarrow}+\overline{\Delta}_{\mathbf{k}}^s\anni{c}{s-\mathbf{k}\downarrow}
\anni{c}{s\mathbf{k}\uparrow}\right)\,,
\label{eqHsmod}
\end{eqnarray}
with 
\begin{eqnarray}
\tilde{\xi}_{\mathbf{k}\sigma}^s&=&\xi_{\mathbf{k}}^s-
\frac{\lambda^2|t^\sigma(\mathbf{k})|^2\,(\xi_{\mathbf{k}}^n+\xi_{\mathbf{k}}^s)}{(\xi_{\mathbf{k}}^n)^2-
(E_{\mathbf{k}}^s)^2}\,,
\label{eqsemod}
\\
\Delta_{\mathbf{k}}^s&=&\Delta_{\mathbf{k}}\cdot\left(1-\frac{\lambda^2}{2}\cdot\frac{|t^\uparrow(\mathbf{k})|^2+|t^\downarrow(\mathbf{k})|^2}{(\xi_{\mathbf{k}}^n)^2-(E_{\mathbf{k}}^s)^2}\right)\,,
\label{eqsdmod}
\end{eqnarray}
being the renormalised kinetic energy and pairing function in the superconductor, and
\begin{eqnarray}
\widehat{\cal H}_n(t)&=&\sum_{\mathbf{k},\sigma}\,\tilde{\xi}_{\mathbf{k}\sigma}^n\,\crea{c}{n\mathbf{k}\sigma}
\anni{c}{n\mathbf{k}\sigma}
\nonumber\\&&\mbox{}
+\sum_{\mathbf{k}}\left(\Delta_{\mathbf{k}}^n\,e^{-2ie\Phi(t)/\hbar}\,\crea{c}{n\mathbf{k}\uparrow}
\crea{c}{n-\mathbf{k}\downarrow}+\overline{\Delta}_{\mathbf{k}}^n\,e^{2ie\Phi(t)/\hbar}\,\anni{c}{n-\mathbf{k}\downarrow}
\anni{c}{n\mathbf{k}\uparrow}\right)\,,
\label{eqHnmod}
\end{eqnarray}
with 
\begin{eqnarray}
\tilde{\xi}_{\mathbf{k}\sigma}^n&=&\xi_{\mathbf{k}}^n+
\frac{\lambda^2|t^\sigma(\mathbf{k})|^2\,(\xi_{\mathbf{k}}^n+\xi_{\mathbf{k}}^s)}{(\xi_{\mathbf{k}}^n)^2-
(E_{\mathbf{k}}^s)^2}\,,
\label{eqnemod}
\\
\Delta_{\mathbf{k}}^n&=&-\frac{\lambda^2}{2}
\cdot\frac{\Delta_{\mathbf{k}}(\,|t^\uparrow(\mathbf{k})|^2+|t^\downarrow(\mathbf{k})|^2\,)}{(\xi_{\mathbf{k}}^n)^2-(E_{\mathbf{k}}^s)^2}\,,
\label{eqndmod}
\end{eqnarray}
being the renormalised kinetic energy and induced pairing function in the normal-metal, due to the proximity effect.
However, such an induced pairing function is multiplied by a time-dependent phase, due to the applied 
voltage across the junction. If one assumes that the tunnelling square amplitudes are independent of 
spin, i.e. $|t^\uparrow(\mathbf{k})|^2=|t^\downarrow(\mathbf{k})|^2$, the modified kinetic energies are also independent of the spin one can, introducing in each $((\mathbf{k},\uparrow),(-\mathbf{k},\downarrow))$ subspace the operators,
\begin{eqnarray}
\widehat{\sigma}_{\mathbf{k}}^z&=&\crea{c}{n\mathbf{k}\uparrow}
\anni{c}{n\mathbf{k}\uparrow}+\crea{c}{n-\mathbf{k}\downarrow}
\anni{c}{n-\mathbf{k}\downarrow}-1\,,
\label{eqsz}
\\
\widehat{\sigma}_{\mathbf{k}}^+&=&\crea{c}{n\mathbf{k}\uparrow}
\crea{c}{n-\mathbf{k}\downarrow}\,,
\label{eqs+}
\\
\widehat{\sigma}_{\mathbf{k}}^-&=&\anni{c}{n-\mathbf{k}\downarrow}
\anni{c}{n\mathbf{k}\uparrow}\,,
\label{eqs-}
\end{eqnarray}
whose algebra is isomorphic to the spin $1/2$ algebra, write $\widehat{\cal H}_n(t)$, up
to a constant factor, as
\begin{equation}
\widehat{\cal H}_n(t)=\sum_{\mathbf{k}}\,\left(\tilde{\xi}_{\mathbf{k}}^n\,
\widehat{\sigma}_{\mathbf{k}}^z
+\Delta_{\mathbf{k}}^n\,e^{-2ie\Phi(t)/\hbar}\,\widehat{\sigma}_{\mathbf{k}}^+
+\overline{\Delta}_{\mathbf{k}}^n\,e^{2ie\Phi(t)/\hbar}\,\widehat{\sigma}_{\mathbf{k}}^-\right)\,.
\label{eqHnmod1}
\end{equation}
We thus conclude that the Hamiltonian describing the normal-metal is equivalent, in each 
$\left((\vecg{k},\uparrow),(-\vecg{k},\downarrow)\right)$ subspace, to the
Hamiltonian of a two-level system under the action of a circularly polarised field. Moreover,
since $\widehat{\cal H}_s$ is time-independent and commutes with $\widehat{\cal H}_n(t)$,
it does not contribute to \eqref{charNIS4}. We can thus write this equation as
\begin{equation}
{\cal G}_{\beta}(\upsilon,\tau)=Tr\,\left[\,\widehat{\mathscr{U}}^\dagger(\tau,0)\,
e^{i\widehat{\cal H}_n(\tau)\upsilon}\,
\widehat{\mathscr{U}}(\tau,0)\,e^{-i\widehat{\cal H}_n(0)\upsilon}\,\widehat{\rho}_n(\beta)\,\right]\,,
\label{charNIS5}
\end{equation}
where 
\begin{equation}
\widehat{\mathscr{U}}(t,0)=T\,\exp\left[-\frac{i}{\hbar} \int_{0}^{t}\,du\,
\widehat{\cal H}_n(u)\,\right]\,,
\label{eqUtransf}
\end{equation}
and $\widehat{\rho}_n(\beta)=e^{-\beta\widehat{\cal H}_n(0)}/{\cal Z}_{n}(\beta)$,
with ${\cal Z}_{n}(\beta)=Tr(e^{-\beta\widehat{\cal H}_n(0)})$. Therefore, we have shown
that there exists an equivalence, in what concerns the calculation of the work characteristic-function in the
adiabatic limit, between the dynamics of the NIS junction and that of an assembly of independent two-level
systems, subjected to a circularly polarised field.

In \ref{secgenfun}, we will consider the calculation of the work characteristic-function
of the latter system.

\section{The work characteristic-function of a two-level system in contact with a thermal bath}
\label{secgenfun}
We now wish to compute \eqref{charNIS5} where $\widehat{\cal H}_n(t)$ is given by \eqref{eqHnmod1}.
Since the different $\mathbf{k}$ modes are independent, we can restrict the calculation
of the said function to that of a single mode. Furthermore, we will drop the $\mathbf{k}$ 
label, as no confusion can arise. 

The derivation of \ref{secMap} shows that one can reduce the dynamics of the system under the applied voltage protocol 
to the dynamics within the low lying energy states of the normal metal if one assumes that the transition time $\tau'\gg \hbar/\Delta$, see eq.
\eqref{eqHnmod1}. We will now further assume that $\Phi(t)\approx\hbar\omega t/2$, where $\omega=2eV_{\text{max}}/\hbar$.
Such an approximation, if it were exact, would actually imply that the voltage was being quenched. For it to be valid, one must have 
$eV_{\mbox{max}}\tau'/\hbar\ll 1$, i.e. $\tau'\ll\hbar/\Delta_n$, as $eV_{\mbox{max}}\approx \Delta_n$. 
Therefore, the time evolution of the system is only adiabatic with respect to the 
larger energy gap, being diabatic \cite{fnoteadiabatic} with respect to the much smaller energy gap,
i.e. one must have $\hbar/\Delta \ll \tau'\ll \hbar/\Delta^n\ll \tau$. Therefore, for a clear separation
between the two time-scales associated with the dynamics of the system to exist, one must have $t^\sigma(\mathbf{k})\ll \Delta$,
as stated above.

We will change somewhat the notation 
with regard to the previous section, so as to keep the result obtained as general as possible,
rather than identifying it solely with the dynamics of the NIS junction. In the new notation,
the initial Hamiltonian is given by
\begin{equation}
\widehat{\cal H}_n(0)=\frac{h}{2}\,\widehat{\sigma}^z+\frac{\Gamma}{2}\,\widehat{\sigma}^x\,,
\label{eqH0}
\end{equation}
where in the NIS junction context $h=2\tilde{\xi}^n$ and 
$\Gamma=2\Delta^n$, and where, without loss of generality, 
$\Delta^n$ can be chosen to be a real-number, as one can always
fix the arbitrary superconducting phase $\varphi$ to be zero. 

The quantities $h$ and $\Gamma$ can be viewed as the components of a 
constant (pseudo) magnetic field applied to the two-level system. 
Since the system is initially in equilibrium with a thermal bath,
the system's partition function is given by
${\cal Z}_{n}(\beta)=2\cosh\left(\beta\sqrt{\Gamma^2+h^2}/2\right)$.

The time-dependent Hamiltonian $\widehat{\cal H}_n(t)$ can be written as
\begin{equation}
\widehat{\cal H}_n(t)=\frac{h}{2}\,\widehat{\sigma}^z+\frac{\Gamma}{2}\left(\,\cos(\omega t)
\,\widehat{\sigma}^x+\sin(\omega t)\,\widehat{\sigma}^y\,\right)\,.
\label{eqHt}
\end{equation}
The dynamics of the two-level system is such that the constant applied field is 
substituted by a  circularly polarised field in $[0,\tau]$, i.e. by a Rabi 
dynamics during that interval. After the application of the protocol,
the circularly polarised field is again replaced by a time-independent field
and the system is described by the constant Hamiltonian $\widehat{H}_n(\tau)$. 
In the process, the applied field has been rotated by an angle $\theta=\omega\tau$ 
around the $z$ axis.

It is well known that for the Rabi dynamics, one can write the time-evolution operator 
$\widehat{\mathscr{U}}(\tau,0)$ in \eqref{charNIS5} as $\widehat{\mathscr{U}}(\tau,0)=
\widehat{\cal R}_\tau\,\widehat{\mathscr{U}}_\tau$, with 
$\widehat{\cal R}_\tau=e^{-i\omega\tau\widehat{\sigma}^z/2}$ representing a pure rotation around 
the $z$ axis and $\widehat{\mathscr{U}}_\tau=e^{-i\widehat{\cal H}'\tau/\hbar}$, 
where $\widehat{\cal H}'$ is a time-independent pseudo-Hamiltonian, which depends on $\omega$, and is given by
\begin{equation}
\widehat{\cal H}'= \frac{1}{2}(h-\hbar\omega)\,\widehat{\sigma}^z+\frac{\Gamma}{2}\,\widehat{\sigma}^x \,.
\label{eqHp}
\end{equation}
Substituting this expression for $\widehat{\mathscr{U}}(\tau,0)$ in \eqref{charNIS5} and noting that 
$\widehat{\cal R}_{\tau}^{\dagger}\, e^{i\widehat{\cal H}_n(\tau) u}\,
\widehat{\cal R}_\tau=
e^{i \widehat{\cal R}_{\tau}^{\dagger}\widehat{\cal H}_n(\tau) \widehat{R}_\tau u}=e^{i\widehat{\cal H}_n(0) u}$, 
since $\widehat{\cal R}_{\tau}^{\dagger}\widehat{\cal H}_n(\tau) \widehat{\cal R}_\tau=\widehat{\cal H}_n(0)$, 
one obtains for the characteristic-function the result
\begin{equation}
{\cal G}_{\beta}(\upsilon,\tau)=\frac{1}{{\cal Z}_{n}(\beta)}Tr\,\left[e^{i\widehat{\cal H}'\tau/\hbar}
e^{i\widehat{\cal H}_n(0) \upsilon}e^{-i\widehat{\cal H}'\tau/\hbar}e^{-i\widehat{\cal H}_n(0) \upsilon}
e^{-\beta\widehat{\cal H}_n(0)}\right]\,.
\label{eqGb1}
\end{equation}
Performing the trace over the complete set of states that diagonalises $\widehat{\cal H}_n(0)$,
we obtain for ${\cal G}_{\beta}(\upsilon,\tau)$, the expression
\begin{eqnarray}
{\cal G}_{\beta}(\upsilon,\tau)&=&\frac{1}{{\cal Z}_{n}(\beta)}\sum_{\sigma,\sigma'}
\mid\bra{\sigma,\unitvec{n}}e^{i\widehat{\cal H}'\tau/\hbar}\ket{\sigma',\unitvec{n}}\mid^2\nonumber\\
&&\mbox{}\times e^{i(\sigma'-\sigma)\upsilon\sqrt{\Gamma^2+h^2}/2}e^{-\beta\sigma\sqrt{\Gamma^2+h^2}/2}\,,
\label{eqGb2}
\end{eqnarray}
where the unit vector $\unitvec{n}$ refers to the direction of the applied field at $t=0$, 
its components being given by $\widehat{n}_x=\frac{\Gamma}{\sqrt{\Gamma^2+h^2}}$ and 
$\widehat{n}_z=\frac{h}{\sqrt{\Gamma^2+h^2}}$. Moreover, writing $\widehat{\cal H}'=\frac{1}{2}\sqrt{\Gamma^2+(h-\hbar\omega)^2}\,
(\unitvec{n}'\cdot\widehat{\boldsymbol{\sigma}})$, where $\widehat{n}_x'=\frac{\Gamma}{\sqrt{\Gamma^2+(h-\hbar\omega)^2}}$ and 
$\widehat{n}_z'=\frac{h-\hbar\omega}{\sqrt{\Gamma^2+(h-\hbar\omega)^2}}$, we can expand the exponential
$e^{i\widehat{\cal H}'\tau/\hbar}$ as
\begin{eqnarray}
e^{i\widehat{\cal H}'\tau/\hbar}&=&\cos\left(\frac{\tau\sqrt{\Gamma^2+(h-\hbar\omega)^2}}{2\hbar}\right)\mathbbm{1}
\nonumber\\
&&\mbox{}+i\,(\unitvec{n}'\cdot\unitvec{n})\sin\left(\frac{\tau\sqrt{\Gamma^2+(h-\hbar\omega)^2}}{2\hbar}\right)
(\unitvec{n}\cdot\widehat{\boldsymbol{\sigma}})\nonumber\\
&&\mbox{}+i\sin\left(\frac{\tau\sqrt{\Gamma^2+(h-\hbar\omega)^2}}{2\hbar}\right)\,[\unitvec{n}
\times(\unitvec{n}'\times\unitvec{n})]\cdot\widehat{\boldsymbol{\sigma}}\,.
\label{eqexpHp}
\end{eqnarray}
The first two terms of \eqref{eqexpHp} are diagonal in the $\ket{\sigma,\unitvec{n}}$ basis,
whereas the last term is only non-zero when evaluated between two states 
of the $\ket{\sigma,\unitvec{n}}$ basis with opposite spin. We thus obtain that
the square moduli of the matrix elements that appear in eq. \eqref{eqGb2} are given by
\begin{eqnarray}
\mid\bra{\sigma,\unitvec{n}}e^{i\widehat{\cal H}'\tau/\hbar}\ket{\sigma',\unitvec{n}}\mid^2&=&\left[\,
\cos^2\left(\frac{\tau\sqrt{\Gamma^2+(h-\hbar\omega)^2}}{2\hbar}\right)\right.\nonumber\\
&&\mbox{}+\left.(\unitvec{n}'\cdot\unitvec{n})^2
\sin^2\left(\frac{\tau\sqrt{\Gamma^2+(h-\hbar\omega)^2}}{2\hbar}\right)\,\right]\delta_{\sigma,\sigma'}\nonumber\\
&&\mbox{}+\mid\unitvec{n}\times(\unitvec{n}'\times\unitvec{n})\mid^2\nonumber\\
&&\mbox{}\times
\sin^2\left(\frac{\tau\sqrt{\Gamma^2+(h-\hbar\omega)^2}}{2\hbar}\right)\delta_{\sigma,-\sigma'}\,.
\label{smel}
\end{eqnarray}
Taking into account that $\mid\unitvec{n}\times(\unitvec{n}'\times\unitvec{n})\mid^2=1-(\unitvec{n}'\cdot\unitvec{n})^2$
and using the expressions for the components of $\unitvec{n}$ and $\unitvec{n}'$ given above in \eqref{smel}, we finally obtain the expression
\begin{eqnarray}
{\cal G}_{\beta}(\upsilon,\tau)&=&\left[\,
\cos^2\left(\frac{\tau\sqrt{\Gamma^2+(h-\hbar\omega)^2}}{2\hbar}\right)\right.\nonumber\\
&&\mbox{}+\left.\frac{(\Gamma^2+h^2-h\hbar\omega)^2}{
(\Gamma^2+h^2)(\Gamma^2+(h-\hbar\omega)^2)}\cdot
\sin^2\left(\frac{\tau\sqrt{\Gamma^2+(h-\hbar\omega)^2}}{2\hbar}\right)\,\right]\nonumber\\
&&\mbox{}+\frac{\Gamma^2(\hbar\omega)^2}{(\Gamma^2+h^2)(\Gamma^2+(h-\hbar\omega)^2)}\cdot
\sin^2\left(\frac{\tau\sqrt{\Gamma^2+(h-\hbar\omega)^2}}{2\hbar}\right)\nonumber\\
&&\mbox{}\times
\frac{\cosh\left[\,\left(\beta/2+i\upsilon\right)\sqrt{\Gamma^2+h^2}\,\right]}{\cosh\left(\beta\sqrt{\Gamma^2+h^2}/2\right)}\,.
\label{eqGb3}
\end{eqnarray}
It is trivial to check that ${\cal G}_{\beta}(\upsilon,\tau)$, as given by \eqref{eqGb3}, fulfils both
the equality ${\cal G}_{\beta}(0,\tau)=1$ and ${\cal G}_{\beta}(i\beta,\tau)=1$, which provides
a check on the correctness of the result, since these equalities are built into the definition
of  ${\cal G}_{\beta}(\upsilon,\tau)$ by construction, as pointed out above. 

Performing the inverse Fourier transform on ${\cal G}_{\beta}(u,\tau)$, we obtain for
$P_{\beta}(W,\tau)$ the result
\begin{eqnarray}
{P}_{\beta}(W,\tau)&=&\left[
\cos^2\left(\frac{\tau\sqrt{\Gamma^2+(h-\hbar\omega)^2}}{2\hbar}\right)\,+\right.\nonumber\\
&&\left.\frac{(\Gamma^2+h^2-h\hbar\omega)^2}{
(\Gamma^2+h^2)(\Gamma^2+(h-\hbar\omega)^2)}
\sin^2\left(\frac{\tau\sqrt{\Gamma^2+(h-\hbar\omega)^2}}{2\hbar}\right)\right]\delta(W)\nonumber\\
&&\mbox{}+\frac{\Gamma^2(\hbar\omega)^2}{(\Gamma^2+h^2)(\Gamma^2+(h-\hbar\omega)^2)}
\sin^2\left(\frac{\tau\sqrt{\Gamma^2+(h-\hbar\omega)^2}}{2\hbar}\right)\nonumber\\
&&\mbox{}\times
\frac{e^{\beta W/2}}{2\cosh\left(\beta\sqrt{\Gamma^2+h^2}/2\right)}\nonumber\\
&&\cdot\left[\delta\left(W-\sqrt{\Gamma^2+h^2}\right)+
\delta\left(W+\sqrt{\Gamma^2+h^2}\right)\right]\,.
\label{eqPW}
\end{eqnarray}
The form of \eqref{eqPW} is easily interpreted from the two-level structure of the
problem, as either no work is performed on the system if the time-dependent field does
not induce transitions between the levels (first term), or otherwise the time-dependent field
induces a transition between the ground state and the excited state, involving
an amount of work performed on the system equal to $W=\sqrt{\Gamma^2+h^2}$, or
the inverse transition is induced involving a negative amount of work $W=-\sqrt{\Gamma^2+h^2}$
being performed on the system (second term). 
Note that if one sets $W\rightarrow -W$, $\omega\rightarrow -\omega$ and $h\rightarrow -h$,
$\Gamma\rightarrow -\Gamma$ in \eqref{eqPW}, one can directly check
that $P_{\beta}(-W,-\tau)=e^{-\beta W}\,P_{\beta}(W,\tau)$, i.e. the work distribution 
satisfies the Crooks-Tasaki relation. 

The average work performed on the system during the application of the protocol 
can be computed either by differentiating \eqref{eqGb3} with respect to $\upsilon$, 
or directly from \eqref{eqPW}, and is given by
\begin{eqnarray}
\langle W \rangle_\beta &=& \frac{\Gamma^2(\hbar\omega)^2}{\sqrt{\Gamma^2+h^2}(\Gamma^2+(h-\hbar\omega)^2)}\,
\sin^2\left(\frac{\tau\sqrt{\Gamma^2+(h-\hbar\omega)^2}}{2\hbar}\right)\nonumber\\
&&\mbox{}\times\tanh\left(\beta\sqrt{\Gamma^2+h^2}/2\right)\,.
\label{eqavW}
\end{eqnarray}
Since the change in the free energy of the system is zero during 
the transformation, this quantity is equal to the energy dissipated
by the system into the thermal bath during the equilibration process
occurring after $t>\tau$. One can easily check that, according to 
\eqref{eqavW}, $\langle W \rangle_\beta$ is always larger or equal to 
zero in agreement with the second law of thermodynamics.
It has a maximum at resonance,
i.e. if $\omega=(\Gamma^2+h^2)/(\hbar h)$ and if 
$\omega\tau\mid\Gamma\mid/\sqrt{\Gamma^2+h^2}=(2n+1)\pi$,
with $n$ being an integer.
Note that for a single two-level system, one can, for each value of $\tau$, choose $\omega$
such that $\langle W \rangle_\beta=0$, i.e. one takes $\omega$ such that
$\tau\sqrt{\Gamma^2+(h-\hbar\omega)^2}/\hbar=2n\pi$.
The existence of such a minimum can be understood from that fact
that for such choice of $\omega$, $e^{i\widehat{H}'\tau/\hbar}=
\mathbbm{1}$, and thus no transitions between the levels are 
induced by the unitary transformation, see also \eqref{eqGb2}. 
This behaviour regarding the dissipated work in a two-level
system mirrors the corresponding behaviour of the Rabi formula for the 
transition probabilities of such a system under the influence of a 
circularly polarised field.

The mean-square deviation of the work performed during the application of the protocol 
can be computed either by differentiating \eqref{eqGb3} twice 
with respect to $\upsilon$, or directly from \eqref{eqPW}, and is given by
\begin{eqnarray}
\langle (\delta W)^ 2 \rangle_\beta &=& \frac{\Gamma^2(\hbar\omega)^2}{(\Gamma^2+(h-\hbar\omega)^2)}\,\sin^2\left(\frac{\tau\sqrt{\Gamma^2+(h-\hbar\omega)^2}}{2\hbar}\right)
\times\nonumber\\
&&\left[1-\frac{\Gamma^2(\hbar\omega)^2}{(\Gamma^2+h^2)(\Gamma^2+(h-\hbar\omega)^2)}\,
\sin^2\left(\frac{\tau\sqrt{\Gamma^2+(h-\hbar\omega)^2}}{2\hbar}\right)\right.\nonumber\\
&&\mbox{}\times\left.
\tanh\left(\beta\sqrt{\Gamma^2+h^2}/2\right)\right]\,.
\label{eqavdW}
\end{eqnarray}

In Sect.~\ref{secDW}, we use \eqref{eqGb3} to compute the work dissipated by
the NIS junction due to the application of the voltage protocol. Since it is
also of interest, albeit not for the solution of the NIS problem, we leave
to \ref{apB} the calculation of the work characteristic-function for 
an isolated two-level system due to the application of a circularly polarised field
to such a system.

\section{The generating function of an isolated two-level system}
\label{apB}
One can use the results obtained in \ref{secgenfun} to compute the characteristic
function or the work distribution function for an isolated system, described by a micro-canonical
ensemble, which undergoes a transformation that is analogous to the one described in that
section, i.e. the system is initially isolated and is coupled to the circularly polarised field 
at $t=0$, being decoupled from it at $t=\tau$. Such characteristic 
function is given by an inverse Laplace transform \cite{Talkner:2008} that 
involves ${\cal G}_{\beta}(\upsilon,\tau)$ and the partition function ${\cal Z}(\beta)=
2\cosh\left(\beta\sqrt{\Gamma^2+h^2}/2\right)$
\begin{equation}
{\cal G}_{E}(\upsilon,\tau)\,\omega_0(E)=\int_{C}\,\frac{d\beta}{2\pi i}\,e^{\beta E}\,
{\cal G}_{\beta}(\upsilon,\tau)\,{\cal Z}(\beta)\,,
\label{eqGE}
\end{equation}
where $E$ is the energy of the system and $\omega_0(E)=\delta\left(E-\sqrt{\Gamma^2+h^2}/2\right)+
\delta\left(E+\sqrt{\Gamma^2+h^2}/2\right)$ is the density of states of the system at $t=0$. The contour
$C$ is the inverse Laplace transform contour from $c-i\infty$ and $c+i\infty$, with $c$ chosen such
that all the singularities of the integrand are located to the left of $c$. In our case, $c=0$. 

Note that these results cannot be generalised to an assembly of two-level systems, since
the work characteristic-function ${\cal G}_{\beta}(\upsilon,\tau)$ is in this case the product of
characteristic-functions for the individual systems.

The integral can be readily performed and after factoring the term $\omega_0(E)$ out, we obtain for 
${\cal G}_{E}(\upsilon,\tau)$ the result
\begin{eqnarray}
{\cal G}_{E}(\upsilon,\tau)&=&\left[
\cos^2\left(\frac{\tau\sqrt{\Gamma^2+(h-\hbar\omega)^2}}{2\hbar}\right)+\right.
\\
&&\left.\frac{(\Gamma^2+h^2-h\hbar\omega)^2}{
(\Gamma^2+h^2)(\Gamma^2+(h-\hbar\omega)^2)}
\sin^2\left(\frac{\tau\sqrt{\Gamma^2+(h-\hbar\omega)^2}}{2\hbar}\right)\right]+\nonumber\\
&&\frac{\Gamma^2(\hbar\omega)^2}{(\Gamma^2+h^2)(\Gamma^2+(h-\hbar\omega)^2)}\,
\sin^2\left(\frac{\tau\sqrt{\Gamma^2+(h-\hbar\omega)^2}}{2\hbar}\right)\,
e^{-2i\upsilon E}\,.\nonumber
\label{eqGE1}
\end{eqnarray}
From this expression, one can obtain, as above, by inverse Fourier transformation of ${\cal G}_{E}(\upsilon,\tau)$,
the work function distribution for an isolated two-level system. This is given by
\begin{eqnarray}
{P}_{E}(W,\tau)&=&\left[
\cos^2\left(\frac{\tau\sqrt{\Gamma^2+(h-\hbar\omega)^2}}{2\hbar}\right)\right.+\nonumber\\
&&\left.\frac{(\Gamma^2+h^2-h\hbar\omega)^2}{
(\Gamma^2+h^2)(\Gamma^2+(h-\hbar\omega)^2)}\,
\sin^2\left(\frac{\tau\sqrt{\Gamma^2+(h-\hbar\omega)^2}}{2\hbar}\right)\,\right]\delta(W)\nonumber\\
&&\mbox{}+\frac{\Gamma^2(\hbar\omega)^2}{(\Gamma^2+h^2)(\Gamma^2+(h-\hbar\omega)^2)}\nonumber\\
&&\mbox{}\times\sin^2\left(\frac{\tau\sqrt{\Gamma^2+(h-\hbar\omega)^2}}{2\hbar}\right)\,\delta\left(W+2E\right)\,.
\label{eqPWE}
\end{eqnarray}
This result can be easily interpreted if one again notes that either no transitions occur between the two levels
and in this case the work performed is zero (first term), or otherwise the work performed is $-2E$ where $E$ is the energy of the initial level (second term). Also, note that $P_{E+W}(-W,-\tau)=P_{E}(W,\tau)$, which is the version of the Crooks-Tasaki relation appropriate for isolated systems in which the 
expressions for the micro-canonical density of states corresponding to the initial and to the final Hamiltonian are identical, since the spectrum of the 
system does not change under application of the work protocol \cite{Talkner:2008}. 

We can also compute the average work dissipated in the transformation,
using the expression for ${\cal G}_{E}(\upsilon,\tau)$ or that for ${P}_{E}(W,\tau)$. This is given by
\begin{equation}
\langle W \rangle_E = -\frac{2E\,\Gamma^2(\hbar\omega)^2}{(\Gamma^2+h^2)(\Gamma^2+(h-\hbar\omega)^2)}\,
\sin^2\left(\frac{\tau\sqrt{\Gamma^2+(h-\hbar\omega)^2}}{2\hbar}\right)\,.
\label{eqavWE}
\end{equation}
Again, one can minimise or maximise this quantity by an appropriate choice of the value of $\omega$. This result
may be relevant in the context of quantum phase-shift gates \cite{Martinis:2002} where the transformation can be implemented 
at finite frequency without generation of heat.

Finally, we can also compute the mean-square deviation of the work performed during the application of the protocol,
using the expression for ${\cal G}_{E}(\upsilon,\tau)$ or that for ${P}_{E}(W,\tau)$. This quantity is given by
\begin{eqnarray}
\langle (\delta W)^2 \rangle_E &=&\frac{4E^2\Gamma^2(\hbar\omega)^2}{(\Gamma^2+h^2)(\Gamma^2+(h-\hbar\omega)^2)}
\sin^2\left(\frac{\tau\sqrt{\Gamma^2+(h-\hbar\omega)^2}}{2\hbar}\right)
\cdot\\
&&\left[1-\frac{\Gamma^2(\hbar\omega)^2}{(\Gamma^2+h^2)(\Gamma^2+(h-\hbar\omega)^2)}
\sin^2\left(\frac{\tau\sqrt{\Gamma^2+(h-\hbar\omega)^2}}{2\hbar}\right)\right].\nonumber
\label{eqavdWE}
\end{eqnarray}
\section{tunnelling matrix element extraction from experiment}
\label{secApTun}
The power dissipated per atom in a NIS junction in the steady state, when submitted to a constant voltage, is given by \cite{Bardeen:1961,Cohen:1962,Ambegaokar:1963,Ambegaokar:1963E,Wolf:2012}
\begin{eqnarray}
{\cal P}&=&
\frac{2\pi eV}{\hbar {\cal N}_{at}}
\sum_{\mathbf{k}\sigma}\,|t^\sigma(\mathbf{k})|^2(f(\xi^n_{\mathbf{k}}-eV)-f(\xi^n_{\mathbf{k}}))
\left[\left|u_\mathbf{k}\right|^2\delta(\xi^n_{\mathbf{k}}-E^s_{\mathbf{k}}-eV)\right.
\nonumber\\
&&\mbox{}
+\left.
\left|v_\mathbf{k}\right|^2\delta(\xi^n_{\mathbf{k}}+E^s_{\mathbf{k}}-eV)\right],
\label{eqPfin}
\end{eqnarray}
where $f(\epsilon)=1/(e^{\beta\epsilon}+1)$ is the Fermi function and where 
\begin{equation}
\left\{
\mbox{
\begin{tabular}{l}   
$|u_{\mathbf{k}}|^2=\frac{1}{2}\left(1+\frac{\xi^s_\mathbf{k}}{E^s_\mathbf{k}}\right)$\\
\\
$|v_{\mathbf{k}}|^2=\frac{1}{2}\left(1-\frac{\xi^s_\mathbf{k}}{E^s_\mathbf{k}}\right)$
\end{tabular}
}\right.\,,
\label{eqBoguv}
\end{equation}
are the squares of the occupation factors in the superconductor.
Using the density of states per atom (and per spin) in the normal metal $\rho(\xi)$ introduced above, we can write this equation as
\begin{equation}
{\cal P}=
\frac{4\pi eV}{\hbar}\int_{-\infty}^{+\infty}\,d\xi\,
\frac{|t(\xi)|^2\rho(\xi)}{\sqrt{\xi^2+\Delta^2(\xi)}}(f(\xi-eV)-f(\xi))
\gamma(\xi),
\label{eqPfin2}
\end{equation}
where the function $\gamma(\xi)$ is given by
\begin{eqnarray}
\gamma(\xi)&=&(\sqrt{\xi^2+\Delta^2(\xi)}+eV/2)\,\delta(\xi-eV-\sqrt{\xi^2+\Delta^2(\xi)})
\nonumber\\
&&\mbox{}+
(\sqrt{\xi^2+\Delta^2(\xi)}-eV/2)\,\delta(\xi-eV+\sqrt{\xi^2+\Delta^2(\xi)})\,.
\label{eqgamma}
\end{eqnarray}

Assuming that the gap in the superconductor is constant within the Debye window of frequencies, one can perform
the above integral, where only the second delta function in the definition of $\gamma(\xi)$ gives a contribution.
We obtain
\begin{equation}
{\cal P}=\frac{2\pi\,|t(\xi_0)|^2\,\Delta^2\,\rho(\xi_0)\sinh(e|V|/2k_BT)}{\hbar\,e|V|[\,\cosh(e|V|/2k_BT)+
\cosh(\Delta^2/2e|V|k_BT)\,]}\,,
\label{eqPsc}
\end{equation}
where $\xi_0=eV/2-\Delta^2/2eV$. This implies that in first-order perturbation theory there is no power dissipated, and hence no current flowing for voltages well below the gap. Note, however, that eq. \eqref{eqavW2} corresponds to a second-order calculation, since $\Delta^n\propto |t|^2$.

Note that in the model described by eq. \eqref{eqHMc}, 
the tunnelling process conserves momentum in the plane of the films, and hence the junction is not ohmic 
when both these films are in the normal state. However, one can measure the power dissipated  when $e|V|=\Delta$ in the superconducting
state of the aluminium film. We obtain from \eqref{eqPsc}
\begin{equation}
{\cal P}=
\frac{\pi}{\hbar}\,|t(0)|^2\rho(0)\Delta\tanh(\Delta/2k_BT)\,.
\label{eqPscT0}
\end{equation}
Therefore, a measurement of ${\cal P}$ at $e|V|=\Delta$, as a function of the temperature, allows the extraction of $|t(0)|^2$. Note
that $\Delta (T)$ is itself a function of the temperature, see \cite{Tinkham:2004}.

Note that an expression of the tunnelling matrix element in terms of parameters of the system such as the thickness of the insulating layer
(see comment on section \ref{secHm}) always requires a model of the potential barrier, see e.g. \cite{Simmons:1963}. The above method avoids the need to resort to such modelling.
\section{Techniques for  measurement of heat released in the NIS junction}
\label{appIM}
Several techniques can be used to measure the small heat release that was computed in section \ref{secIso}. The most straightforward technique
would be the use of standard calorimetry \cite{White:2002,Pobell:2007}. In such a case, if the conversion of the work
performed on the film, into heat, occurs in a short time frame after the end of the work protocol, there is an increase $\Delta T$ in the temperature of the normal metal film given by
\begin{equation}
\Delta T=\frac{\langle W\rangle_\beta}{n c_v}\,,
\label{eqDT}
\end{equation}
where $n$ is the number of moles of material contained in the probe, 
and $c_v$ the material's molar heat capacity. If we use $\langle W\rangle_\beta\approx 0.7\,\units{n} \units{J}/\units{g}$,
as computed in section \ref{secIso}, we obtain a variation of temperature $\Delta T=1.28\times 10^{-5} \,\units{K}$,
likely to be too small to be measured.

A possible alternative technique would be the measurement of the released heat at constant temperature by coupling the system 
to an external probe that absorbs a quantity of heat, $\Delta Q=\langle W\rangle_\beta$, isothermally. This transformation would be analogous to the 
isothermal expansion phase of a Stirling engine \cite{Callen:1960}. If an infinitesimal amount of heat $\delta Q$ is released in the NIS junction during the equilibration process, after application of the protocol (as stated in section \ref{secIso}, the equilibration time is of the order of the electron-phonon relaxation time $\tau_{\text{pe}}$), we have from the second law of thermodynamics, that $\delta Q/T=dS_{\text{pr}}+dS_{\text{HB}}$, where $dS_{\text{pr}}$ is the variation of the entropy of the probe (which would be the elastic degrees of freedom of the normal metal film) and $dS_{\text{HB}}$ is the variation of entropy of the  heat bath (surrounding helium bath). Since the temperature, volume and number of atoms of the bath do not change, $dS_{\text{HB}}=0$.
Thus, we have
\begin{equation}
\delta Q/T=dS_{\text{pr}}=\left.\derp{S}{\mathscr{V}}\right|_{T}\,d\mathscr{V}=\left.\derp{P}{T}\right|_{\mathscr{V}}\,d\mathscr{V}=-\frac{\derp{\mathscr{V}}{T}|_P}{\derp{\mathscr{V}}{P}|_{T}}\,d\mathscr{V}=
\frac{\alpha}{k_T}\,d\mathscr{V}\,,
\label{diffS}
\end{equation}
where we have applied the Maxwell relation on going from the second to the third equality and the implicit function theorem on going
from the third to the fourth equality in the equation above, and where $\kappa_T$ is the isothermal compressibility of the probe and $\alpha$ its thermal expansion coefficient, as follows from equilibrium thermodynamics. The reasoning above is analogous to the one made when discussing 
the measurement of the
magnetocaloric effect from magnetisation data in isothermal conditions, see Ref. \cite{Pecharsky:1999}.  Note that the stresses in the film and the 
pressure of the fluid that makes up the heat bath
do not need to be equal and moreover, that the mechanical element that controls the expansion of the film would only move in time 
scales much larger than $\tau_{pe}$, necessary for the equilibration of the elastic degrees of freedom of the film.

Assuming that the thermodynamic quantities that
enter eq. \eqref{diffS} are independent of the volume for the small variations measured, we obtain, integrating this equation
\begin{equation}
\Delta {\mathscr{V}}=\frac{\kappa_T}{\alpha T}\,\Delta Q\,.
\label{eqVS}
\end{equation}
Note that in this case the variation of volume is inversely proportional
to the thermal expansion parameter, which implies that the smallness
of this quantity in aluminium at low temperatures will actually amplify the 
effect to be measured.

Therefore, the (average) relative variation of the volume of the probe is given by 
\begin{equation}
\frac{\Delta {\mathscr{V}}}{\mathscr{V}}=\frac{\langle W\rangle_\beta}{\gamma n c_v T}\,,
\label{eqVW}
\end{equation}
where $\gamma=\frac{\alpha v_m}{c_v k_T}$ is the Gr\"uneisen parameter
of the material that constitutes the probe and $v_m$ is its molar volume. 
For aluminium, $\gamma\approx 1.7$ \cite{Gauster:1971} and $c_v\approx 1.5\,\units{mJ}\,
\units{mol}^{-1}\,\units{K}^{-1}$ \cite{Pobell:2007}. 

The measurement of the relative variation of the probe's volume in isothermal conditions 
thus gives direct experimental access to the work produced during the prescribed protocol.
A histogram of the distribution of data obtained for multiple realizations of the 
protocol will allow the reconstruction of the full work distribution.

If we use $\langle W\rangle_\beta\approx 0.7\,\units{n} \units{J}/\units{g}$, 
we obtain a relative deviation $\frac{\Delta {\mathscr{V}}}{\mathscr{V}}\approx 7.5\times10^{-6}$,
on average, which should be measurable using, e.g. capacitive methods or SQUID dilatometers, as
such a relative deviation is in the limit of precision of capacitive methods \cite{White:2002,Pobell:2007}, 
provided that other sources of heat dissipation can be minimized. 

\section*{References}

\begin{thebibliography}{100}
\bibitem{fnoteadiabatic}
For a full adiabatic evolution of the system, one must have, from the adiabatic
  theorem, $\tau'\gg \hbar eV_{\text{max}}/(\Delta^n)^2 \approx
  \hbar/\Delta^n$, since $\Delta^n$ is the smallest gap in the system and $
  eV_{\text{max}}$ is the rate of change of the Hamiltonian with respect to the
  dimensionless time $t/\tau'$.
\end{thebibliography}

\end{document}